\documentclass[preprint]{aastex}

\usepackage{paralist}
\usepackage{amsmath}
\usepackage{epstopdf}
\usepackage{graphicx}

\shorttitle{An ancient system with five sub-Earth-size planets}                               
\shortauthors{Campante et al.}

\begin{document}

\title{An ancient extrasolar system with five sub-Earth-size planets}

\author{T.~L.~Campante\altaffilmark{1,2}}
\email{campante@bison.ph.bham.ac.uk}
\and
\author{
    T.~Barclay\altaffilmark{3,4},
    J.~J.~Swift\altaffilmark{5},
    D.~Huber\altaffilmark{3,6,7},
    V.~Zh.~Adibekyan\altaffilmark{8,9},
    W.~Cochran\altaffilmark{10},
    C.~J.~Burke\altaffilmark{6,3},
    H.~Isaacson\altaffilmark{11},
    E.~V.~Quintana\altaffilmark{6,3},
    G.~R.~Davies\altaffilmark{1,2},
    V.~Silva Aguirre\altaffilmark{2},
    D.~Ragozzine\altaffilmark{12},
    R.~Riddle\altaffilmark{13},
    C.~Baranec\altaffilmark{14},
    S.~Basu\altaffilmark{15},
    W.~J.~Chaplin\altaffilmark{1,2},
    J.~Christensen-Dalsgaard\altaffilmark{2},
    T.~S.~Metcalfe\altaffilmark{16,2},
    T.~R.~Bedding\altaffilmark{7,2}, 
    R.~Handberg\altaffilmark{1,2},
    D.~Stello\altaffilmark{7,2},
    J.~M.~Brewer\altaffilmark{17},
    S.~Hekker\altaffilmark{18,2},
    C.~Karoff\altaffilmark{2,19},
    R.~Kolbl\altaffilmark{11},
    N.~M.~Law\altaffilmark{20},
    M.~Lundkvist\altaffilmark{2},
    A.~Miglio\altaffilmark{1,2},
    J.~F.~Rowe\altaffilmark{3,6},
    N.~C.~Santos\altaffilmark{8,9,21},
    C.~Van Laerhoven\altaffilmark{22},
    T.~Arentoft\altaffilmark{2},
    Y.~P.~Elsworth\altaffilmark{1,2},
    D.~A.~Fischer\altaffilmark{17},
    S.~D.~Kawaler\altaffilmark{23},
    H.~Kjeldsen\altaffilmark{2},
    M.~N.~Lund\altaffilmark{2},
    G.~W.~Marcy\altaffilmark{11},
    S.~G.~Sousa\altaffilmark{8,9,21},
    A.~Sozzetti\altaffilmark{24},
    T.~R.~White\altaffilmark{25}
}

\altaffiltext{1}{School of Physics and Astronomy, University of Birmingham, Edgbaston, Birmingham, B15 2TT, UK}
   
\altaffiltext{2}{Stellar Astrophysics Centre (SAC), Department of Physics and Astronomy, Aarhus University, Ny Munkegade 120, DK-8000 Aarhus C, Denmark}   

\altaffiltext{3}{NASA Ames Research Center, Moffett Field, CA 94035, USA}
   
\altaffiltext{4}{Bay Area Environmental Research Institute, 596 1st Street West, Sonoma, CA 95476, USA}

\altaffiltext{5}{Department of Astronomy and Department of Planetary Science, California Institute of Technology, MC 249-17, Pasadena, CA 91125, USA}

\altaffiltext{6}{SETI Institute, 189 Bernardo Avenue \#100, Mountain View, CA 94043, USA}

\altaffiltext{7}{Sydney Institute for Astronomy, School of Physics, University of Sydney, Sydney, Australia}

\altaffiltext{8}{Centro de Astrof\'isica, Universidade do Porto, Rua das Estrelas, 4150-762 Porto, Portugal}

\altaffiltext{9}{Instituto de Astrof\'isica e Ci\^encias do Espa\c{c}o, Universidade do Porto, Rua das Estrelas, 4150-762 Porto, Portugal}

\altaffiltext{10}{Department of Astronomy and McDonald Observatory, The University of Texas at Austin, TX 78712-1205, USA}

\altaffiltext{11}{Astronomy Department, University of California, Berkeley, CA 94720, USA}
   
\altaffiltext{12}{Department of Physics and Space Sciences, Florida Institute of Technology, 150 West University Boulevard, Melbourne, FL 32901, USA}
   
\altaffiltext{13}{Division of Physics, Mathematics, and Astronomy, California Institute of Technology, Pasadena, CA 91125, USA}   

\altaffiltext{14}{Institute for Astronomy, University of Hawai`i at M\=anoa, Hilo, HI 96720-2700, USA}

\altaffiltext{15}{Department of Astronomy, Yale University, New Haven, CT 06520, USA}

\altaffiltext{16}{Space Science Institute, Boulder, CO 80301, USA}
   
\altaffiltext{17}{Department of Physics, Yale University, New Haven, CT 06511, USA}

\altaffiltext{18}{Max Planck Institute for Solar System Research, D-37077 G\"ottingen, Germany}

\altaffiltext{19}{Department of Geoscience, Aarhus University, H{\o}egh-Guldbergs Gade 2, DK-8000 Aarhus C, Denmark}

\altaffiltext{20}{Department of Physics and Astronomy, University of North Carolina at Chapel Hill, Chapel Hill, NC 27599-3255, USA}

\altaffiltext{21}{Departamento de F\'isica e Astronomia, Faculdade de Ci\^encias, Universidade do Porto, Rua do Campo Alegre, 4169-007 Porto, Portugal}

\altaffiltext{22}{Department of Planetary Sciences, University of Arizona, 1629 E University Blvd., Tucson, AZ 85721, USA}

\altaffiltext{23}{Department of Physics and Astronomy, Iowa State University, Ames, IA 50011, USA}

\altaffiltext{24}{INAF -- Osservatorio Astrofisico di Torino, Via Osservatorio 20, I-10025 Pino Torinese, Italy}

\altaffiltext{25}{Institut fur Astrophysik, Georg-August-Universit\"at G\"ottingen, Friedrich-Hund-Platz 1, 37077 G\"ottingen, Germany}

\begin{abstract}
The chemical composition of stars hosting small exoplanets (with radii less than four Earth radii) appears to be more diverse than that of gas-giant hosts, which tend to be metal-rich. This implies that small, including Earth-size, planets may have readily formed at earlier epochs in the Universe's history when metals were more scarce. We report {\it Kepler} spacecraft observations of Kepler-444, a metal-poor Sun-like star from the old population of the Galactic thick disk and the host to a compact system of five transiting planets with sizes between those of Mercury and Venus. We validate this system as a true five-planet system orbiting the target star and provide a detailed characterization of its planetary and orbital parameters based on an analysis of the transit photometry. Kepler-444 is the densest star with detected solar-like oscillations. We use asteroseismology to directly measure a precise age of $11.2\pm1.0\:{\rm Gyr}$ for the host star, indicating that Kepler-444 formed when the Universe was less than $20\,\%$ of its current age and making it the oldest known system of terrestrial-size planets. We thus show that Earth-size planets have formed throughout most of the Universe's $13.8$-billion-year history, leaving open the possibility for the existence of ancient life in the Galaxy. The age of Kepler-444 not only suggests that thick-disk stars were among the hosts to the first Galactic planets, but may also help to pinpoint the beginning of the {\it era of planet formation}. 
\end{abstract}

\keywords{Galaxy: disk --- planetary systems --- stars: individual (HIP~94931) --- stars: late-type --- stars: oscillations --- techniques: photometric}

\section{Introduction}\label{sec:intro}
Transit and radial velocity surveys have found an increasing number of planets with low mass and/or radius orbiting other suns \citep{BoruckiCandidates2,MayorTwins,Dumusque}. Some of these planets may even be found in the habitable zones around their parent stars \citep{Pepe,Quintana14}. The precision of transit measurements, in combination with mass determinations from radial velocity measurements and planet interior models, have also allowed the determination of the bulk composition of several planets \citep{Leger,Howard,Pepe13}. For the most favorable cases, exquisite measurements further allowed detection of both the emitted (infrared) and reflected (optical) light of exoplanets, as well as of specific atmospheric lines \citep{Brogi12,Rodler}. These measurements are providing a first insight into the physics of exoplanet atmospheres.

The NASA {\it Kepler} mission was designed to use the transit method to find Earth-like planets in and near the habitable zones of late-type main-sequence stars \citep{KeplerMission,KeplerDesign}. {\it Kepler} has so far successfully yielded over $4{,}000$ exoplanet candidates, of which approximately $40\,\%$ are in multiple-planet systems \citep{BoruckiCandidates1,BoruckiCandidates2,BatalhaCandidates3,BurkeCandidates4}. The recent announcement of a wealth of new multiple-planet systems \citep{Lissauer14,Rowe14} has raised the number of exoplanets confirmed by {\it Kepler} to nearly one thousand, a sample including planets as diverse as hot-Jupiters, super-Earths, or even circumbinary planets.

Transit observations being an indirect detection technique are, however, only capable of providing planetary properties relative to the properties of the parent star. Asteroseismology can play an important role in the characterization of host stars and thus allow for inferences on the properties of their planetary companions. In particular, the information contained in solar-like oscillations allows fundamental stellar properties (i.e., density, surface gravity, mass, radius, and age) to be precisely determined \citep{CunhaReview,ChaplinMiglio}. Prior to the advent of {\it Kepler}, solar-like oscillations had been detected in only a few tens of main-sequence and subgiant stars using ground-based high-precision spectroscopy \citep{Brown91,BouchyCarrier,Arentoft08} or ultra-high-precision, wide-field photometry from the {\it CoRoT} space telescope \citep{Appour08,MichelSci}. Photometry from {\it Kepler} has since led to an order of magnitude increase in the number of such stars with confirmed oscillations \citep{ChaplinSci,VernerPipelines}.

Early seismic studies of exoplanet-host stars were conducted using ground-based \citep{Bouchy05,Vauclair08} and {\it CoRoT} data \citep{Gaulme10,Ballot11}. The first application of asteroseismology to known exoplanet hosts in the {\it Kepler} field \citep{JCDExo} has been followed by a series of planet discoveries where asteroseismology was used to constrain the properties of the host star \citep{Kepler10,TrES2,Kepler-37,Kepler69,Kepler22,Kepler36,Kepler21,ChaplinObliquities,Kepler68,Misalignment,Ballard,Kepler410A}. \citet{HuberEnsembleKOI} presented the first systematic study of {\it Kepler} planet-candidate host stars using asteroseismology. More recently, \citet{CampanteDetect} provided lower limits on the surface gravities of planet-candidate host stars from the non-detection of solar-like oscillations.

In this work we report {\it Kepler} spacecraft observations of Kepler-444 (also known as \object[HIP 94931]{HIP~94931}, KIC~6278762, KOI-3158, and LHS~3450), a seismic K dwarf from the old Galactic thick disk and the host to a compact system of five transiting, sub-Earth-size planets. Transit-like signals, indicative of five planets, were detected by {\it Kepler} over the course of four years of nearly continuous observations of Kepler-444 \citep{Tenenbaum}. The five planets in this system have been identified as planet candidates in the NASA Exoplanet Archive\footnote{\url{http://exoplanetarchive.ipac.caltech.edu/}} \citep{NASAExoArchive} based on the examination of {\it Kepler} pixel-level and light-curve data. Kepler-444 is a cool main-sequence star of spectral type K0V \citep{Roman55,Eggen56,Wilson62}. It is a high-proper-motion star \citep[$\mu_\alpha \cos\delta\!=\!98.94\pm0.80\:\rm{mas\,yr^{-1}}$ in right ascension and $\mu_\delta\!=\!-632.49\pm0.85\:\rm{mas\,yr^{-1}}$ in declination;][]{Hipparcos}, with an annual proper motion in excess of $0.5\:{\rm arcsec}$. A value of $-121.19\pm0.11\:{\rm km\,s^{-1}}$ for the radial velocity is given in \citet{Latham02}, while \citet{GenCop} report a value of $-121.9\pm0.1\:{\rm km\,s^{-1}}$. Its reported {\it Hipparcos} parallax \citep[$\pi\!=\!28.03\pm0.82\:{\rm mas}$;][]{Hipparcos} makes it one of the closest confirmed {\it Kepler} planetary systems, at a distance $d\!=\!35.7\pm1.1\:{\rm pc}$ in the constellation Lyra, commensurate with the distances to Kepler-3 \citep[a bright K dwarf with a transiting hot-Neptune at $d\!=\!36.4\pm1.3\:{\rm pc}$;][]{Bakos10} and Kepler-42 \citep[an M dwarf with three transiting sub-Earth-size planets at $d\!=\!38.7\pm6.3\:{\rm pc}$;][]{Muirhead12}. It is among the brightest {\it Kepler} planetary hosts with a {\it Kepler}-band magnitude $Kp\!=\!8.717$ \citep{KIC} and apparent magnitude $V\!=\!8.86$. Kepler-444 is also iron-poor, as described by a number of independent studies providing atmospheric parameter estimates based on high-resolution spectroscopy \citep{Peterson80,Tomkin99,Soubiran08,Sozzetti09,Ramya12}. From these we determined a mean iron abundance of $[{\rm Fe}/{\rm H}]\!=\!-0.69\pm0.09\:{\rm dex}$ ($0.20\:{\rm dex}$ scatter).

The rest of the paper is organized as follows. In Sect.~\ref{sec:ground} we present the spectroscopic analysis and high-resolution imaging that were part of a ground-based follow-up. This is followed in Sect.~\ref{sec:origin} by an analysis of the parent star's chemical properties and kinematics in establishing the system's thick-disk membership. A comprehensive asteroseismic analysis is conducted in Sect.~\ref{sec:seismic} where emphasis is placed on the estimation of a precise stellar age. Sect.~\ref{sec:planetary} provides a validation and characterization of the planetary system. Finally, a discussion of the results and conclusions are presented in Sect.~\ref{sec:discussion}.

\section{Ground-based follow-up}\label{sec:ground}

\subsection{Spectroscopic analysis}\label{sec:spec}
\begin{figure}[!t]
\centering
\includegraphics[width=\linewidth]{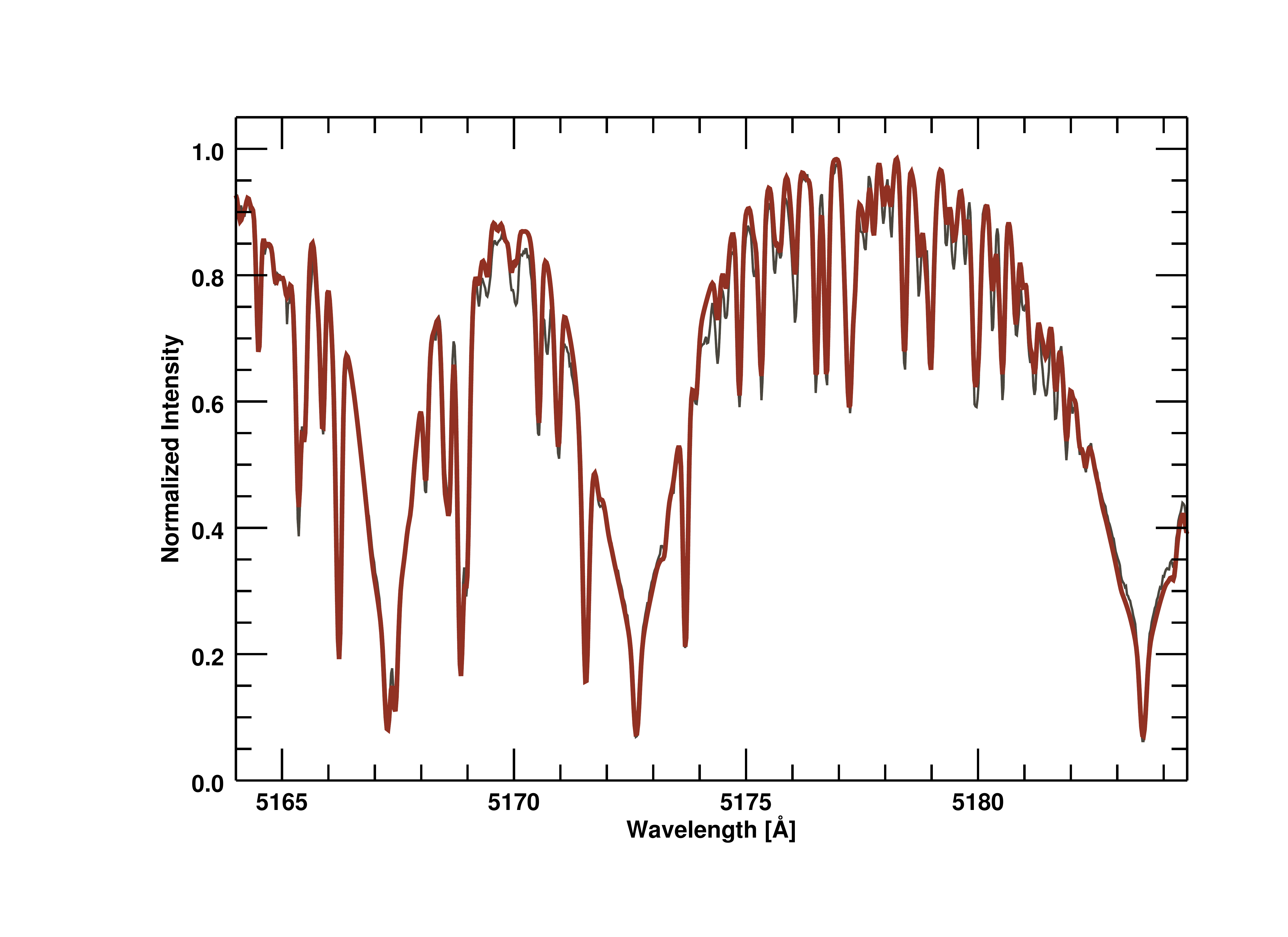}
\caption{\small Observed and synthetic spectra in the wavelength region occupied by the \ion{Mg}{1} b triplet lines. The spectral fitting procedure covered $160\:{\rm\AA}$ with most of the coverage in the range $6000$--$6200\:{\rm\AA}$, while having also included the region around the gravity-sensitive \ion{Mg}{1} b triplet lines. The synthetic spectrum (thick line) provides a good fit to the \ion{Mg}{1} b line wings. The combined effect of rotational line broadening and macroturbulence takes a value of $\sim\!2.2\:{\rm km\,s^{-1}}$, an indication that this star is a fairly slow rotator.\label{fig:spec}}
\end{figure}

To provide values of the effective temperature, $T_{\rm eff}$, surface gravity, $\log g$, and elemental abundances for Kepler-444, we acquired a high-resolution spectrum with the Keck I telescope and HIRES spectrograph \citep{HIRES} on 2012 July 2. Use of the standard setup and reduction of the California Planet Search \citep{Johnson10} resulted in a spectrum with resolving power $R\!\approx\!60{,}000$ and signal-to-noise ratio ($S/N$) of 200 per pixel. The local-thermodynamic-equilibrium (LTE) analysis package SME \citep[Spectroscopy Made Easy;][]{Valenti96} was then used following the setup and procedures of \citet{Valenti05} and \citet{Valenti09} to determine the atmospheric parameters and elemental abundances given in Table \ref{tb:spec} (see also Fig.~\ref{fig:spec}). Contributions of $59\:{\rm K}$ in $T_{\rm eff}$ and $0.062\:{\rm dex}$ in metallicity, $[{\rm Fe}/{\rm H}]$, were added in quadrature to the formal uncertainties to account for systematic differences between spectroscopic methods \citep{Torres12}. These values of $T_{\rm eff}$ and $[{\rm Fe}/{\rm H}]$ will later be used as input to the asteroseismic analysis (Sect.~\ref{sec:starprop}). The tabulated value for $[{\rm Fe}/{\rm H}]$ is consistent with the mean literature value within the reported scatter.

We averaged the individual relative abundances of silicon and titanium to obtain a measure of the relative abundance of $\alpha$ elements, $[\alpha/{\rm Fe}]$. This yields $[\alpha/{\rm Fe}]\!=\!0.26\pm0.07\:{\rm dex}$ after accounting for systematics \citep{Torres12}. This is in fair agreement with a proxy value of $0.17\pm0.10\:{\rm dex}$ from Str\"omgren photometry \citep{Casagrande11}. Kepler-444 is iron-poor and moderately $\alpha$-enhanced, following the well-established Galactic trend of increasing $\alpha$-enhancement with decreasing metallicity \citep{AllerGreenstein,Wallerstein}.

\subsection{On the detection of a bound M-dwarf pair}\label{sec:pair}
During spectroscopic observations of Kepler-444 with the Keck telescope, a fainter companion $1.8\:{\rm arcsec}$ to the west was visually detected on the HIRES guide camera. Given {\it Kepler}'s detector scale of $3.98\:{\rm arcsec\,pixel^{-1}}$, these stars are unresolved in {\it Kepler} observations. Average seeing on Mauna Kea of $1.0\:{\rm arcsec}$ allows separation of the primary and secondary for all acquired spectra. The two components of the Kepler-444 system are co-moving as implied by their systemic radial velocity, which varies by less than $3\:{\rm km\,s^{-1}}$. This system has been reported as being a possible S-type binary by \citet{Lillo-Box}. With a {\it Hipparcos} determined distance of $35.7\:{\rm pc}$ and visual separation of $1.8\:{\rm arcsec}$, we estimate the orbital period of the secondary around the primary to be $\sim\!430\:{\rm yr}$, with some uncertainty due to the projected angle of the orbital plane relative to the line of sight and the unknown eccentricity.

Unlike for the primary, an LTE analysis of the spectrum of the secondary is inappropriate, since a cross-correlation with the template spectrum of the M dwarf GL~699 (Barnard's star) showed two peaks. This is an indication that the secondary is in fact composed of two M dwarfs, meaning that the present system is a hierarchical triple system. Using a newly developed routine \citep{Kolbl}, we cross-correlated the M-dwarf binary composite spectrum with a library of 700 well-understood spectra collected by the California Planet Search. Upon finding the best match to the library (in a $\chi^2$ sense), the primary spectrum was subtracted from the composite spectrum, and the residuals were computed. The brighter of the two M dwarfs closely matches a star with $T_{\rm eff}\!=\!3464\:{\rm K}$ and $\log g\!=\!5.0\:{\rm dex}$. Returned uncertainties are $200\:{\rm K}$ and $0.2\:{\rm dex}$, respectively. A precise measurement of their relative brightnesses is made difficult due to the imperfect subtraction of the primary from the composite spectrum. The residuals best match a star in the temperature range from 3500 to $4000\:{\rm K}$. The two stars being equidistant, we would expect the secondary to be slightly cooler than the primary. While we are unable to directly measure the surface gravity of the fainter M dwarf, the association with the two other components in the system and the expected cool temperature suggest a value of $\log g\!\sim\!5\:{\rm dex}$.

\subsection{High-resolution adaptive optics imaging}\label{sec:AO}
Kepler-444 was observed with the Robo-AO laser-adaptive-optics system \citep{Baranec13} at the Palomar Observatory 60-inch telescope on 2013 July 21 to look for contaminating sources within the {\it Kepler} aperture. We used a long-pass filter with a 600-nm cut-on (LP600) to more closely approximate the {\it Kepler} bandpass while maintaining diffraction-limited resolution \citep{Law}. The observation consisted of a sequence of full-frame-transfer detector readouts at the maximum frame rate of $8.6\:{\rm Hz}$ for a total of $90\:{\rm s}$ of integration time. The individual images were then combined using ex post facto shift-and-add processing taking Kepler-444 as the tip-tilt star with $100\,\%$ frame selection \citep{Baranec14}. 

\begin{figure}[!t]
\centering
\includegraphics[width=\linewidth]{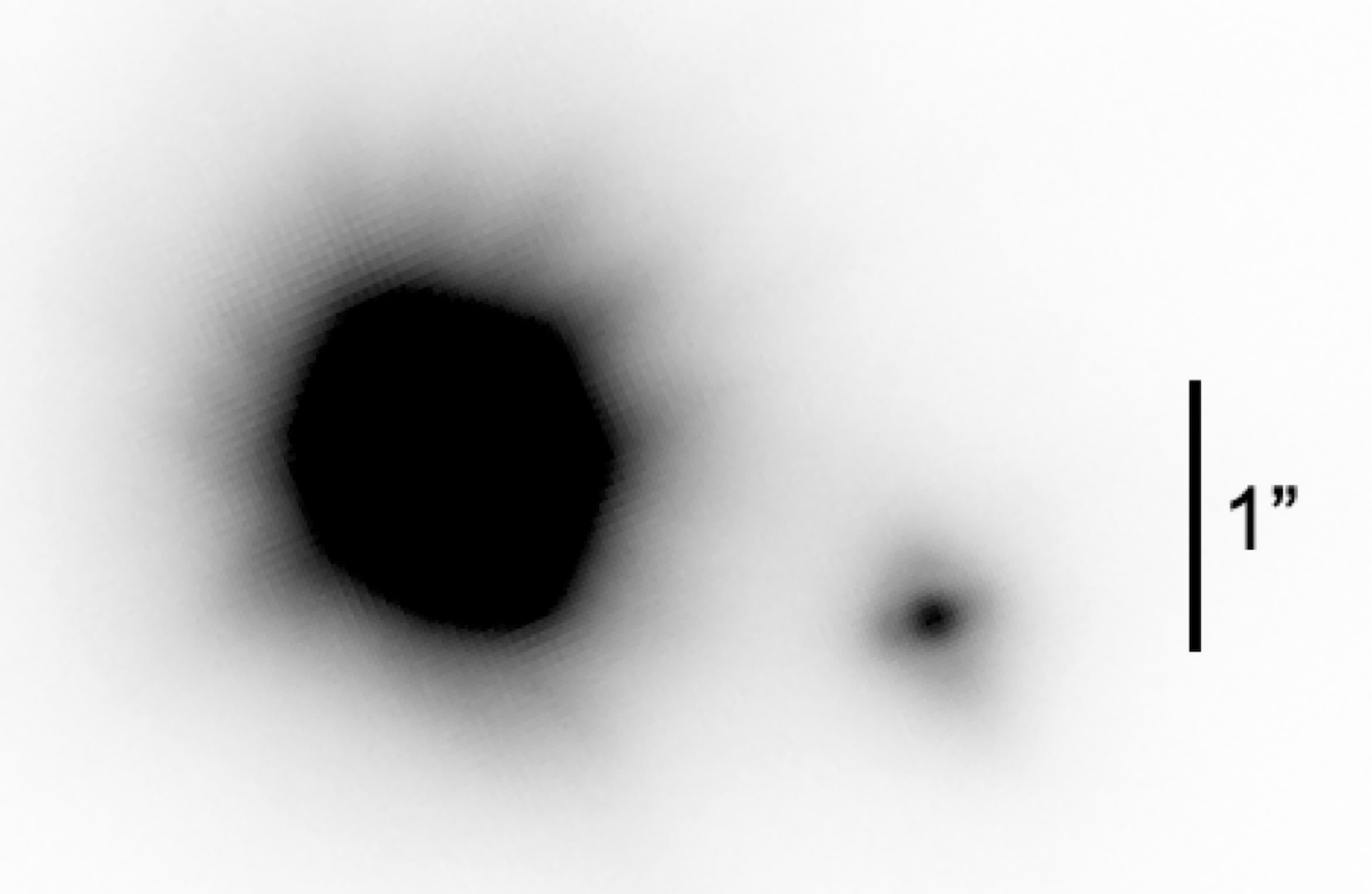}
\caption{\small Kepler-444 resolved into two components by Robo-AO. This linearly-scaled image was obtained using a long-pass filter (LP600) that approximately matches the {\it Kepler} bandpass at the redder visible wavelengths. The magnitude difference between the two components was measured to be $\Delta m_{\rm LP600}\!=\!3.51\pm0.02$. The secondary lies roughly to the west (to the right in the image), at a position angle of $251\pm2\:{\rm deg}$ with respect to the primary.\label{fig:AO}}
\end{figure}

We detected that Kepler-444 is a binary (no split is seen in the secondary with this technique) with a separation of $1.87\pm0. 03\:{\rm arcsec}$ (Fig.~\ref{fig:AO}). We used simple aperture photometry to calculate the flux ratio of the primary and secondary components. We measured the total flux centered on each component and subtracted off an equivalent aperture on the opposite side of the companion to approximate subtraction of both the stellar halo and sky background. Multiple aperture sizes were used to generate estimates of the systematic errors, and we found consistent flux ratios when using apertures from $0.2$ to $0.4\:{\rm arcsec}$. The magnitude difference between the two components was measured to be $\Delta m_{\rm LP600}\!=\!3.51\pm0.02$. Detection of a previously unknown companion within the {\it Kepler} aperture of a host star will affect the derived radius of any planet candidate transiting that star, since the {\it Kepler} observed transit depth is shallower than the true depth due to dilution \citep{Kepler14b}. From the magnitude difference given above, we estimate a dilution of $3.94\pm0.08\,\%$ in the {\it Kepler} bandpass.

\begin{figure}[!t]
\centering
\includegraphics[width=\linewidth]{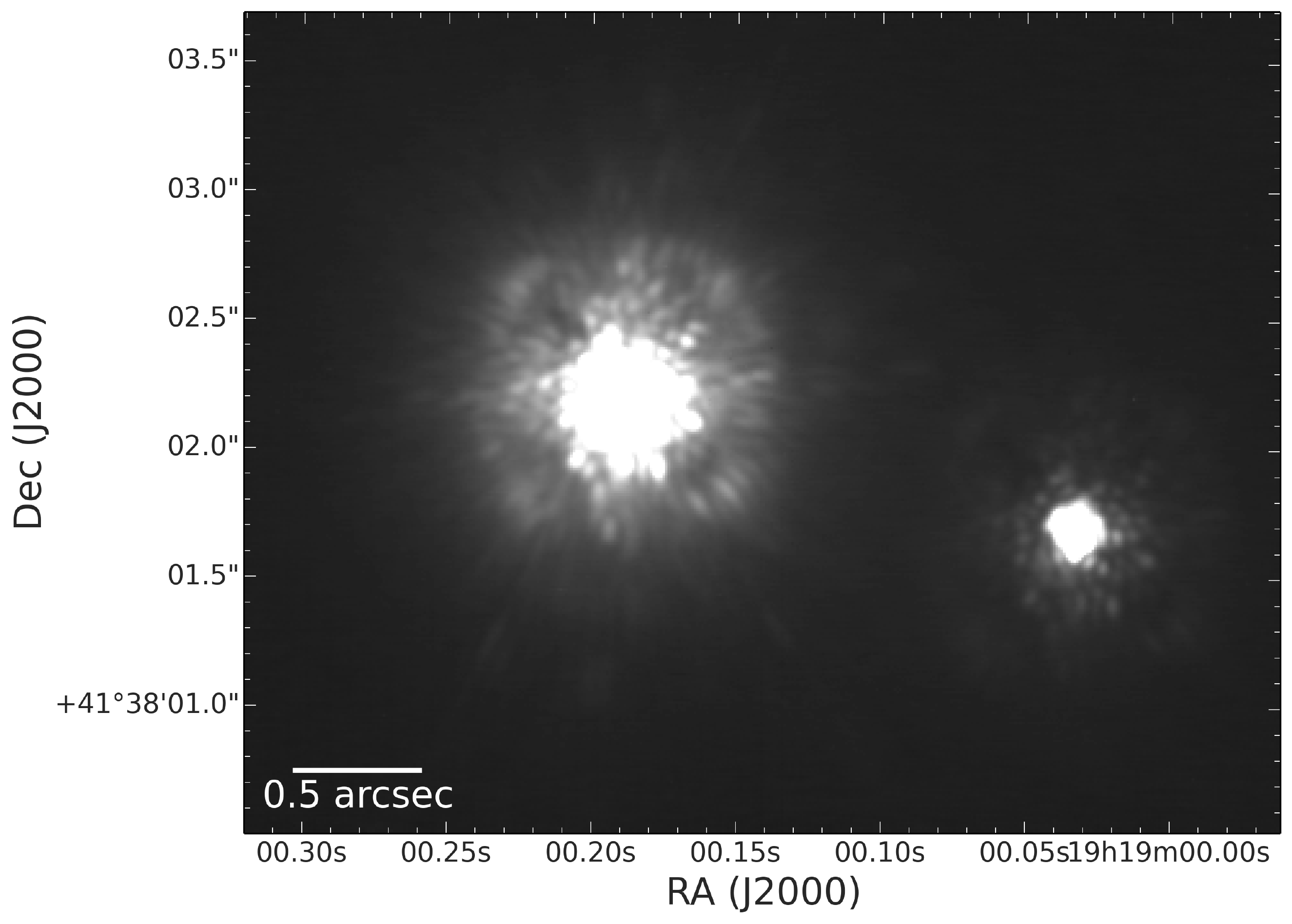}
\caption{\small NIRC2 adaptive optics image of Kepler-444. The image was obtained using the $K^\prime$ filter ($2.124\:{\rm \mu m}$) for a total of $378\:{\rm s}$ of integration time. Declination and right ascension coordinates (J2000.0) are given along the vertical and horizontal axes, respectively.\label{fig:AO2}}
\end{figure}

We further observed Kepler-444 on 2014 August 8 with the NIRC2 instrument mounted on the Keck II 10-meter telescope. We used the $K^\prime$ filter (with central wavelength $\lambda_{\rm c}\!\sim\!2.124\:{\rm \mu m}$). Despite very poor weather conditions on the night the observations were made, it was still possible to collect data through sporadic breaks in the clouds owing to the target's brightness. There are 36 usable frames in total. Nine of the frames each consist of twelve coadds of $0.8$-second integrations, while the remaining frames each consist of eighteen coadds of $0.6$-second integrations. Therefore, the total integration time of usable data amounts to $378\:{\rm s}$. The FWHM of the stellar point spread function in the combined image is of $0.07\:{\rm arcsec}$ (Fig.~\ref{fig:AO2}). We see no evidence for any additional stars in the system besides the primary and the secondary. In other words, we are once more unable to resolve the secondary M-dwarf pair.

\section{System's provenance and its place in the Galaxy}\label{sec:origin}

\subsection{On the $\alpha$-element overabundance of metal-poor exoplanet-host stars}\label{sec:chemical}
The correlation between the occurrence of giant planets and the metallicity of host stars is now well established, with metal-rich stars being more likely to harbor gas giants \citep{Gonzalez-97,Santos-01,Santos-04,Fischer-05,Johnson-10,Mayor-11,Petigura-11,Sousa-11,Mortier-13}, which in turn lends support to the model that giant planets form by concurrent accretion of solids and gas \citep{Pollack96}. However, this correlation is weakened as one moves toward Neptune-size planets \citep{Ghezzi-10,Mayor-11,Sousa-11}, ultimately vanishing as we enter the regime of Earth-size planets \citep{Buchhave-12,Buchhave14}. Based on the spectroscopic metallicities of the host stars of 226 {\it Kepler} exoplanet candidates, it has been shown \citep{Buchhave-12} that the metallicity distribution of stars harboring small planets (i.e., with radii less than $4\,R_\earth$) is rather flat and covers a wide range of metallicities. This could mean that the process of formation of small planets is less constrained than that of the formation of large planets, with rocky planets likely starting to form at an earlier epoch than gas giants \citep{Fischer-12}.

Most of the studies aimed at clarifying whether or not exoplanet-host stars differ from stars without planets in their content of individual heavy elements showed no significant difference between the two populations \citep{Fischer-05,Takeda-07,Neves-09,DelgadoMena-10,GonHernandez-13}. A number of studies have, nonetheless, reported possible enrichment of some species in host stars \citep{Gonzalez-01,Gilli-06,Robinson-06,Brugamyer-11,Kang-11,Adibekyan-12a}. Based on a chemical abundance analysis of a large sample of F, G, and K dwarfs from the HARPS GTO (Guaranteed Time of Observation) planet search program, it has been shown \citep{Adibekyan-12c,Adibekyan-12a} that the vast majority of host stars are overabundant in $\alpha$ elements in the low-metallicity regime, i.e., for $[{\rm Fe}/{\rm H}]\!<\!-0.3\:{\rm dex}$. This result could be confirmed in a follow-up to that work \citep{Adibekyan-12b}, now using a combination of the HARPS GTO sample and a subset of the \citet{Buchhave-12} {\it Kepler} sample. This was to be expected, since there are other fairly abundant elements (e.g., the $\alpha$ elements Mg and Si) with condensation temperatures comparable to iron \citep{Lodders-03}, whose contributions to the composition of dust and rocky material in planet-forming regions are very important. Most of the planets in this iron-poor regime were found to be super-Earth-size or Neptune-size planets. Consequently, and although small planets can be found in an iron-poor regime, their host stars will most likely be overabundant in $\alpha$ elements. This result implies that the early formation of rocky planets could have started in the Galactic thick disk, where chemical conditions are more favorable compared to the thin disk \citep{Reddy06}. Similarly favorable conditions seem to be associated with a fraction of the halo stellar population, namely, the so-called high-$\alpha$ stars \citep{Nissen10}.

\subsection{Determining the system's thick-disk membership}\label{sec:thick}
\begin{figure}[!t]
\centering
\includegraphics[width=\linewidth]{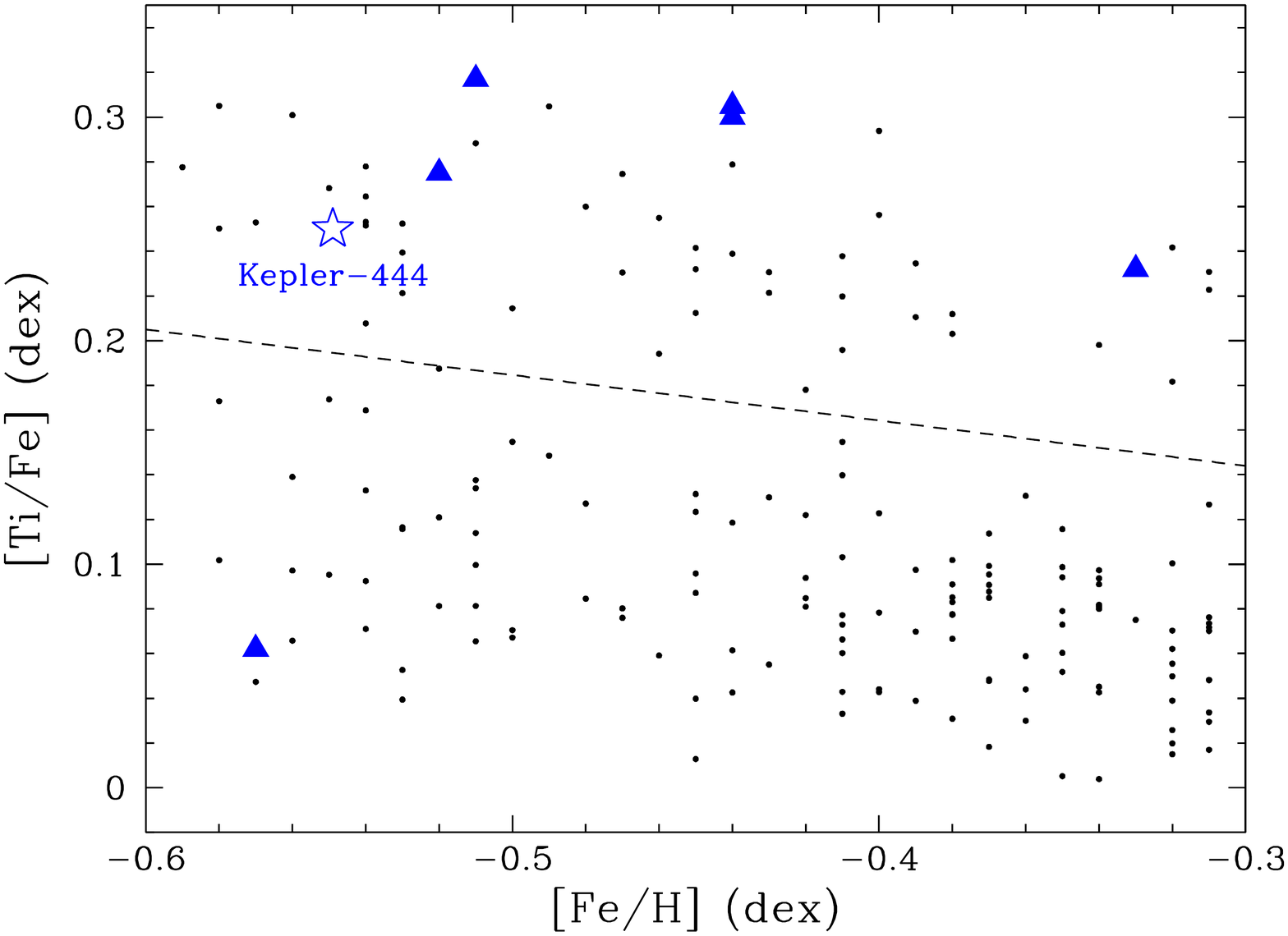}
\caption{\small $[{\rm Ti}/{\rm Fe}]$ abundance ratio versus $[{\rm Fe}/{\rm H}]$ in the low-metallicity regime. Kepler-444 is denoted by a blue star. Blue triangles represent the {\it Kepler} planet-candidate host stars of \citet{Adibekyan-12b} with available titanium abundances. Black dots correspond to stars without planetary companions observed in the context of the HARPS GTO planet search program. The dashed line marks the fiducial chemical separation between the thin and thick disks \citep[below and above the line, respectively;][]{Adibekyan-12b} based solely on the HARPS GTO sample.\label{fig:ti_fe}}
\end{figure}

Figure \ref{fig:ti_fe} shows the $[{\rm Ti}/{\rm Fe}]$ abundance ratio (taken as a proxy of $[\alpha/{\rm Fe}]$) versus $[{\rm Fe}/{\rm H}]$ (in the range $-0.6\!<\![{\rm Fe}/{\rm H}]\!<\!-0.3\:{\rm dex}$) for Kepler-444 and the {\it Kepler} planet-candidate host stars of \citet{Adibekyan-12b} with available titanium abundances. For comparison, stars without planetary companions observed in the context of the HARPS GTO planet search program were also included. The stars are divided into two groups according to their Ti content, with the dashed line marking the fiducial chemical separation between the thin and thick disks \citep{Adibekyan-12b}. Thick-disk stars are overabundant in Ti with respect to this dividing line, whereas thin-disk stars exhibit low Ti content. We decided to use $[{\rm Ti}/{\rm Fe}]$ as a proxy of $[\alpha/{\rm Fe}]$, since this allows for a clear separation between the thin and thick disks in $[{\rm Ti}/{\rm Fe}]$ vs.~$[{\rm Fe}/{\rm H}]$ space \citep{Bensby-03,Bensby14,Neves-09,Adibekyan-12a}. Kepler-444 is then seen to belong to the Galactic thick disk based on its Ti content ($[{\rm Ti}/{\rm Fe}]\!=\!0.250\pm0.083\:{\rm dex}$), while fitting the paradigm of iron-poor host stars being overabundant in $\alpha$ elements \citep{Adibekyan-12b,Adibekyan-12a}. We should note that five out of the six remaining {\it Kepler} planet-candidate host stars are also overabundant in Ti and hence likely thick-disk members.

\begin{figure}[!t]
\centering
\includegraphics[width=\linewidth]{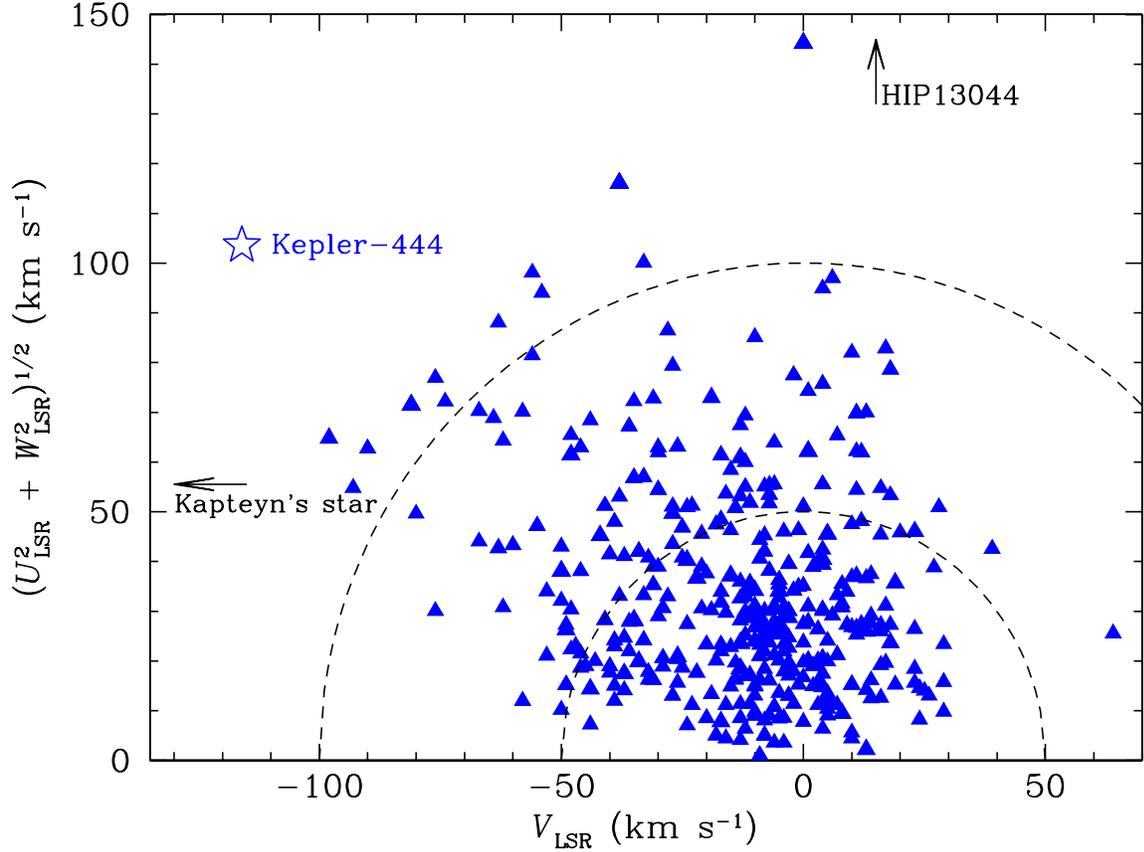}
\caption{\small Toomre diagram showing all exoplanet-host stars in The Extrasolar Planets Encyclopaedia for which it was possible to derive Galactic space velocities ($U_{\rm LSR},V_{\rm LSR},W_{\rm LSR}$) relative to the Local Standard of Rest (LSR). Kepler-444 is denoted by a blue star. HIP~13044 and Kapteyn's star fall outside the plotted range. Dashed lines represent curves of constant peculiar velocity, viz., $\upsilon_{\rm pec}\!=\!50$ and $100\:{\rm km\,s^{-1}}$.\label{fig:toomre}}
\end{figure}

We also examined the kinematics of Kepler-444 to assess the likelihood of it being a member of the Galactic halo. A Toomre diagram is shown in Fig.~\ref{fig:toomre}, where we represent all exoplanet-host stars in The Extrasolar Planets Encyclopaedia\footnote{\url{http://exoplanet.eu/}} \citep{exoplanet.eu} for which it was possible to derive Galactic space velocities (370 systems harboring 490 planets). We followed \citet{Adibekyan-12c} when computing the Galactic velocities ($U_{\rm LSR},V_{\rm LSR},W_{\rm LSR}$) relative to the Local Standard of Rest (LSR) for this sample. Galactic velocity components for Kepler-444 were derived using the radial velocity reported by \citet{GenCop}, along with the {\it Hipparcos} parallax and proper motion \citep{Hipparcos}. We obtained $(U_{\rm LSR},V_{\rm LSR},W_{\rm LSR})\!=\!(67.0\pm2.4,-114.0\pm0.4,-79.0\pm1.6)\:{\rm km\,s^{-1}}$, and thus a peculiar velocity $\upsilon_{\rm pec}\!\equiv\!(U^2_{\rm LSR} + V^2_{\rm LSR} + W^2_{\rm LSR})^{1/2}\!\approx\!154\:{\rm km\,s^{-1}}$. Kepler-444 has the third largest peculiar velocity of all stars depicted in Fig.~\ref{fig:toomre} after HIP~13044 and Kapteyn's star, whose peculiar velocities take the values $\upsilon_{\rm pec}\!\approx\!420\:{\rm km\,s^{-1}}$ and $\upsilon_{\rm pec}\!\approx\!280\:{\rm km\,s^{-1}}$, respectively. HIP~13044 belongs to the Helmi stream and is likely of extragalactic origin. An earlier claim that this horizontal-branch star harbored a giant planet \citep{Setiawan-10} has been recently contested \citep{Jones14}. Therefore, Kepler-444 is to the best of our knowledge the exoplanet-host star with the second largest peculiar velocity after Kapteyn's star, a member of the Galactic halo. We computed the probabilities \citep{Reddy06} that Kepler-444 belongs to the thick disk and to the halo, having adopted both the \citet{Bensby-03} and \citet{Robin-03} population fractions. We found that the star belongs to the thick disk with a probability of $97\,\%$ or $92\,\%$, and to the halo with a probability of $3\,\%$ or $8\,\%$, depending on whether we used the \citeauthor{Bensby-03}~or \citeauthor{Robin-03}~prescription, respectively. In conclusion, we can safely state that Kepler-444 belongs to the Galactic thick disk based on the analysis of both its chemical properties and kinematics.

\subsection{On the possibility of an extragalactic origin}\label{sec:extragalactic}
Kepler-444 is known to belong to the Arcturus stellar stream based on an analysis of the fine structure in the phase space distribution of nearby subdwarfs \citep{AF06}, which revealed the stream as an overdensity in phase space centered at $V_{\rm LSR}\!\sim\!-125\:{\rm km\,s^{-1}}$ and $(U^2_{\rm LSR} + 2V^2_{\rm LSR})^{1/2}\!\sim\!185\:{\rm km\,s^{-1}}$. The Arcturus stellar stream is a moving group from the Galactic thick disk \citep{Eggen71}. It is named after its most illustrious member, Arcturus ($\alpha$ Bo\"otis), the brightest star in the northern celestial hemisphere. 

The origin of the Arcturus stellar stream has been the matter of debate \citep{Klement10}. Several theories have been put forward to explain the origin of stellar streams: \begin{inparaenum}[(i)] \item dissolution of an open cluster; \item dynamical perturbations within the Galaxy; or \item an external origin as a result of the tidal debris from an accreted satellite galaxy. \end{inparaenum} Initially, the Arcturus stream was interpreted as originating from the debris of a disrupted satellite \citep{Navarro04,AF06,Helmi06}. Its angular momentum was thought to be too low to arise from dynamical perturbations induced by the Galactic bar \citep{Navarro04} while coinciding with that of the debris identified by \citet{Gilmore02} lying below and above the Galactic plane, of which the Arcturus stream could very well be the solar-neighborhood extension. The scenario of a dissolved open cluster has been ruled out \citep{Williams09,Ramya12}, since members of the stream were found to be chemically inhomogeneous. The same authors also found that members of the stream are chemically similar to thick-disk field stars, which in turn have abundance properties dissimilar from those of satellite galaxies in the Local Group. These results would then, by exclusion, be compatible with a dynamical origin \citep{Bensby14}. For instance, an explanation for the stream's origin based on dynamical perturbations -- excited by a satellite galaxy -- within the thick disk has been proposed \citep{Minchev09,Gomez}. The very small velocity dispersion found for the Arcturus stream has also been taken as evidence of a dynamical origin \citep{Bovy09}.

\section{Asteroseismic analysis}\label{sec:seismic}
The high-quality photometric data provided by {\it Kepler} are well-suited for conducting asteroseismic studies of stars. In particular, {\it Kepler} short-cadence data \citep[with cadence $\Delta t\!\sim\!1\:{\rm min}$;][]{SCdata} make it possible to investigate solar-like oscillations in main-sequence and subgiant stars, whose dominant periods are of the order of several minutes. Kepler-444 had been observed in short cadence for one month (second monthly segment of Quarter 4) in the context of the {\it Kepler} Asteroseismic Science Consortium \citep[KASC;][]{KASC1,KASC2} during the mission's survey phase. The survey data did not, however, show a detection of solar-like oscillations. Fortunately, this target would later be included in the target proposal for Quarter 6 (Q6), with a main objective being the asteroseismic study of main-sequence stars having temperatures similar to, or cooler than, the Sun. The expectation was that multi-month time series would raise the probability of detections in bright cool dwarfs \citep{CampanteDetect}. These data did show a definitive detection of solar-like oscillations in Kepler-444. Note that the M-dwarf companions are too faint, and their oscillation amplitudes too small, to be detected \citep{ChaplinDetect}. The target has since been observed in short cadence during Q15, Q16, and Q17. In May 2013, the spacecraft lost the second of four gyroscope-like reaction wheels, ending new data collection for the nominal mission \citep{Kepler2}. This coincided with the early stages of data collection during Q17.

\subsection{Global asteroseismic parameters}\label{sec:avgpar}
Solar-like oscillations are predominantly global standing acoustic waves. These so-called p modes -- with the pressure gradient playing the role of the restoring force -- are characterized by being intrinsically damped while simultaneously stochastically excited by near-surface convection \citep{JCDReview}. Therefore, all stars cool enough to harbor an outer convective envelope may be expected to exhibit solar-like oscillations. The oscillation modes are characterized by the radial order $n$ (related to the number of radial nodes), the spherical degree $l$ (specifying the number of nodal surface lines), and the azimuthal order $m$ (with $|m|$ specifying how many of the nodal surface lines are lines of longitude). The observed oscillation modes are typically high-order modes of low spherical degree with the associated power spectrum showing a pattern of peaks with near-regular frequency separations \citep{Vandakurov,Tassoul}. The most prominent separation is the large frequency separation, $\Delta \nu$, between neighboring overtones having the same spherical degree. The large frequency separation essentially scales as $\langle\rho\rangle^{1/2}$ \citep{Ulrich86,Brown94}, where $\langle\rho\rangle\!\propto\!M/R^{3}$ is the mean density of a star with mass $M$ and radius $R$. To second order, the spectrum is also characterized by the small frequency separations $\delta\nu_{02}$ (viz., the frequency spacing between adjacent modes with $l\!=\!0$ and $l\!=\!2$) and $\delta\nu_{01}$ (viz., the amount by which modes with $l\!=\!1$ are offset from the midpoint between the $l\!=\!0$ modes on either side). For main-sequence dwarfs, $\delta\nu_{02}$ and $\delta\nu_{01}$ depend largely on the sound speed gradient in the central regions of the star, gradually decreasing with increasing stellar age \citep{CD84,CD88}. Oscillation mode power is modulated by an envelope that generally assumes a Gaussian shape \citep{Kallinger10}. The frequency at the peak of the power envelope of the oscillations, where the observed modes attain their strongest amplitudes, is referred to as the frequency of maximum oscillation amplitude, $\nu_{\rm max}$. The frequency of maximum oscillation amplitude scales to very good approximation as $g\,T_{\rm eff}^{-1/2}$ \citep{Brown91,KjeldsenBedding95,Belkacem11}. The fact that $\nu_{\rm max}$ mainly depends on the surface gravity, $g$, makes it an indicator of the evolutionary state of a star.

Global asteroseismic parameters, indicative of the overall stellar structure, can be readily obtained using automated analysis methods \citep{VernerPipelines}. We began by phase-clipping all transit signals from the Q6 and Q15 time-series data, noting that the induced gaps in each of the time series -- leading to a duty cycle reduction of about $8\,\%$ -- have a negligible effect on the resulting power spectra for the purpose of computing global asteroseismic parameters. Both time series were then high-pass filtered by applying a quadratic Savitzky--Golay filter \citep{SavitzkyGolay} to remove additional low-frequency power due to stellar activity and instrumental variability. Finally, the corresponding weighted power spectra were computed and then averaged. Working with an average power spectrum is justified by the fact that the two time series are not contiguous, having been collected more than two years apart. Besides, averaging leads to a reduction of the variance in the power spectrum by a factor of two.

\begin{figure}[!t]
\centering
\includegraphics[width=\linewidth]{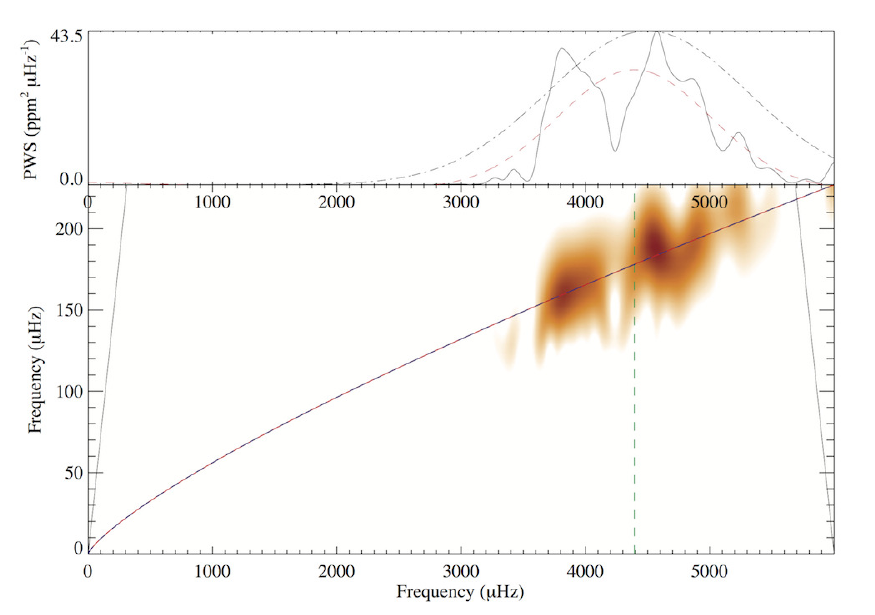}
\caption{\small Wavelet analysis of the oscillation power spectrum. Bottom panel: Wavelet transform using a Morlet wavelet. The transform is affected by two priors. The first is that $\Delta\nu$ and $\nu_{\rm max}$ must be consistent with a simple scaling relation \citep[alternate blue-and-red curve;][]{StelloScaling}. This allows us to reduce the impact of power that lies well away from the scaling relation (by imposing a Gaussian fall-off with a characteristic scale on either side of this curve). The second prior is applied to $\nu_{\rm max}$ and kept wide enough as $\nu_{\rm max}\!=\!4500\!\pm\!800 \: {\rm \mu Hz}$ (represented by a dash-dotted curve in the top panel). The estimate of $\nu_{\rm max}$ returned by this analysis is represented by a vertical dashed line. The slanted solid lines toward the edges of the plotting area delimit the cone of influence. Top panel: Same power wavelet spectrum (PWS) with priors applied, but now integrated over the ordinate frequency (i.e., over all scales). The dashed curve represents a Gaussian fit to the wavelet spectrum.\label{fig:wavelet}}
\end{figure}

The parameters $\Delta \nu$ and $\nu_{\rm max}$ were returned by five automated methods based on the analysis of the above average power spectrum: AAU \citep{AAU,CampantePhD}, KAB \citep{KAB}, OCT \citep{OCT}, SYD \citep{SYD}, and a wavelet-based analysis (Fig.~\ref{fig:wavelet}). To be commensurate with the work of \citet{HuberEnsembleKOI}, we adopted the values for $\Delta \nu$ and $\nu_{\rm max}$ returned by the SYD pipeline, with final uncertainties recalculated by adding in quadrature the formal uncertainty and the standard deviation of the values returned by all contributing methods. The final adopted values are then $\Delta \nu\!=\!179.64\pm0.76\:{\rm \mu Hz}$ and $\nu_{\rm max}\!=\!4538\pm144\:{\rm \mu Hz}$. Note the high precision (better than $1\,\%$) with which $\Delta\nu$ has been measured. Remarkably, Kepler-444 is the star with the largest $\Delta\nu$ ever measured, meaning that Kepler-37 \citep{Kepler-37} has been dethroned as the densest star with detected solar-like oscillations (the mean density of Kepler-444 is estimated in the next section).

A note of caution is in order concerning the computation of $\nu_{\rm max}$. A power spectrum with modes of oscillation necessarily contains structure that repeats itself with some characteristic frequency, $\Delta\nu$. Moreover, the repeated structure is located in some region of frequency space around $\nu_{\rm max}$. By taking a wavelet transform of the power spectrum (Fig.~\ref{fig:wavelet}) we are able to visualize periodic structure at different scales (i.e., in terms of $\Delta\nu$ shown along the ordinate axis in the bottom panel) and at different locations (i.e., in terms of $\nu_{\rm max}$ shown along the abscissa axis in the top and bottom panels). The resulting two-dimensional wavelet transform contains higher power when the scale and location match the observed periodic structure. According to the output of the wavelet analysis, the power envelope of the oscillations appears to be double-humped and not Gaussian-shaped as generally assumed, meaning that $\nu_{\rm max}$ is ill-defined. The higher-frequency hump peaks at a frequency close to the seventh harmonic of the inverse of the long-cadence period \citep{SCdata}. This artifact is, however, too weak -- at least for such a bright {\it Kepler} target -- to produce the observed effect. A double hump has also been observed in the power envelope of $\alpha$ Centauri B \citep{alphaCenB}, yet another K dwarf, an indication that $\nu_{\rm max}$ may become ambiguous for low-mass dwarfs. However, a double hump is not apparent in an alternative power spectrum of Kepler-444 computed for the purpose of conducting detailed frequency modeling (Sect.~\ref{sec:freqmod}). We thus have reasons to believe that the observed double hump in Fig.~\ref{fig:wavelet} is an artifact resulting from the averaging of the power spectra in combination with the finite mode lifetimes. Nevertheless, we note that the value of $\nu_{\rm max}$ is still consistent with $\Delta\nu$ (at the $1\sigma$ level) based on a simple scaling relation \citep{StelloScaling}.

\subsection{Estimation of fundamental stellar properties}\label{sec:starprop}
Fundamental stellar properties can be estimated by comparing global asteroseismic parameters (normally, $\Delta\nu$ and $\nu_{\rm max}$) and complementary spectroscopic observables to the outputs of stellar evolutionary models. To that end, we followed a grid-based approach, whereby observables are matched to well-sampled grids of stellar evolutionary models (or isochrones). A total of five pipeline codes were used coupled to ten stellar evolutionary grids or analysis methodologies. The wide variety of grid-pipeline combinations then implicitly accounts for the impact on the final estimates of using different stellar models -- covering a range of adopted input physics and parameters -- and different analysis methodologies. The determination of stellar properties through a grid-based approach that uses asteroseismic constraints is currently well established \citep{Stello09,Basu10,Basu12,ChaplinSci,Creevey12}. Moreover, the systematic biases involved with grid-based approaches have been the subject of several detailed studies \citep{Gai11,Basu12,Bazot12,Grub12,ChaplinGrid}.

The following grid-based pipeline codes were used:
\begin{enumerate}
\item Asteroseismology Made Easy \citep[AME;][]{AME};
\item Bellaterra Stellar Properties Pipeline \citetext{BeSPP; A.~M.~Serenelli et al., in preparation};
\item Rapid Algorithm for Diameter Identification of Unclassified Stars \citep[RADIUS;][]{Stello09};
\item SEEK \citep{SEEK};
\item Yale--Birmingham \citep[YB;][]{Basu10,Basu12,Gai11}.
\end{enumerate}
The AME pipeline is based on a grid of stellar evolutionary models computed using the Modules for Experiments in Stellar Astrophysics \citep[MESA;][]{Paxton11,Paxton13} code with simple input physics. An early version of the AME pipeline was used in this work. The BeSPP pipeline was run with one of its two stellar evolutionary grids \citep{VSA14}, namely, the grid of BaSTI models \citep[Bag of Stellar Tracks and Isochrones;][]{Pietrinferni04}. Both the RADIUS and SEEK pipelines use grids of models constructed with the Aarhus STellar Evolution Code \citep[ASTEC;][]{ASTEC}, although with different input physics and parameters. The RADIUS pipeline provided two sets of results: one set is based on the properties of the most likely model, while the other set is based on the average properties of a range of acceptable models (i.e., whose parameters lie within $3\sigma$ of the observations). The YB pipeline uses five different stellar evolutionary grids: a grid of models from the Dartmouth group \citep{Dotter08}, a grid of models from the Padova group \citep{Girardi00,Marigo08}, the models comprising the Yonsei--Yale \citep[YY;][]{Demarque04} isochrones, and two grids of models -- named YREC \citep{Gai11} and YREC2 \citep{Basu12} -- constructed with the Yale Rotating stellar Evolution Code \citep[YREC;][]{Demarque08}. \citet{ChaplinGrid} provide an overview of the analysis methodology at work in each of the above pipeline codes and of the adopted input physics and parameters.   

Normally, stellar properties would be estimated using $\{\Delta\nu,\nu_{\rm max},T_{\rm eff},[{\rm Fe}/{\rm H}]\}$ as input. However, given ambiguity in the determination of $\nu_{\rm max}$ for this particular star, we have instead used $\{\Delta\nu,L,T_{\rm eff},[{\rm Fe}/{\rm H}]\}$ as input, where $L$ is the stellar luminosity. The luminosity was estimated from a knowledge of the distance to the star, $d$, and the bolometric flux arriving on Earth, $F_{\rm bol}\!=\!9.2935\!\times\!10^{-9}\:{\rm mW\,m^{-2}}$ \citep[][]{Casagrande11}, which was derived via the InfraRed Flux Method \citep[IRFM;][]{Casagrande10}. A reddening correction can be safely ignored \citep{Casagrande11}, leading to $L/{\rm L}_\sun\!=\!0.37\pm0.03$. There were two exceptions to the aforementioned guideline: AME used $\{\Delta\nu,T_{\rm eff},[{\rm Fe}/{\rm H}]\}$ as input, while SEEK used $\{\Delta\nu,\delta\nu_{02},T_{\rm eff},[{\rm Fe}/{\rm H}]\}$ as input, where an estimate of the small frequency separation, $\delta\nu_{02}$, provided by the KAB automated method was adopted.

We now address the issue of $\alpha$-enhancement, which we take into account in two different ways in our grid-based search. Firstly, one could use preexistent $\alpha$-enhanced grids that closely match the $[\alpha/{\rm Fe}]$ estimate. The main drawback of this approach is that not all of the pipeline codes have access to $\alpha$-enhanced grids. Secondly, one could adopt the prescription of \citet{Salaris93} to mimick $\alpha$-enhanced isochrones, thereby using available nonenhanced grids and a scaled overall metallicity of $[{\rm m}/{\rm H}]\!=\!-0.37\pm0.09\:{\rm dex}$ as input. In the end, we decided to adopt the latter approach. As a sanity check, the YB pipeline was also run using two preexistent $\alpha$-enhanced grids (with $[\alpha/{\rm Fe}]\!=\!0.2$), namely, the Dartmouth and YY grids. For a given grid, estimated stellar properties turned out to be consistent with the ones obtained using the \citeauthor{Salaris93}~prescription, with grid-to-grid systematics being more important.

We provide consolidated values from grid-based modeling for the stellar mass, $M$, radius, $R$, surface gravity, $\log g$, and mean density, $\langle\rho\rangle$, in Table \ref{tb:properties}. To properly account for systematics, these values are given by the median over the contributing grids/pipelines, after a preliminary step that involved the rejection of outliers following Peirce's criterion \citep{Peirce,Gould}. Associated uncertainties are estimated by adding in quadrature the median formal uncertainty and the scatter over the contributing grids/pipelines. The uncertainties in the derived mass, radius, $\log g$, and mean density are $5.7\,\%$, $1.9\,\%$, $0.0095\:{\rm dex}$, and $1.1\,\%$, respectively. These are in very good agreement with the median final uncertainties quoted by \citet{HuberEnsembleKOI}, and those quoted by \citet{ChaplinGrid} for their spectroscopic subset. Unlike surface gravity, density, and radius (and hence luminosity), which are largely model-independent \citep{Lebreton08,Monteiro09}, mass (the same could be said about age) is considerably more sensitive to the input physics and parameters, in particular to the chemical composition. The inflated uncertainty on the latter property is, therefore, a result of the increased grid-to-grid (and pipeline-to-pipeline) scatter \citep{ChaplinGrid}. An iterative procedure is often used to refine the estimates of the spectroscopic parameters by repeating the spectroscopic analysis with $\log g$ fixed at the asteroseismic value, $\log g_{\rm seis}$ \citep{Bruntt12}. Such an iterative procedure was not deemed necessary in the present case, since the asteroseismic and spectroscopic values of $\log g$ were found to be consistent.

\subsubsection{A precise stellar age from asteroseismology}\label{sec:freqmod}
Modeling of the frequencies of individual modes of oscillation, or their combinations, can lead to considerably more precise estimates of the stellar properties, most notably the age \citep{Metcalfe10,Metcalfe12,VSA13}. We thus conducted a detailed frequency modeling in an attempt to precisely estimate the age of the parent star. To this purpose, we used the longest, contiguous short-cadence {\it Kepler} data set available, i.e., by concatenating Q15 and Q16. These data were prepared using an optimized prescription \citep{KASOCpipe}. The time series was corrected by first redefining the pixel masks used for the aperture photometry to include more pixels. Following this step, a combination of moving-median filters were used to remove both long-term trends and short-term anomalies from the time series. The known periods of the five planets were also taken into account in the removal of any periodicities introduced into the light curve. Once this corrected time series had been produced, a weighted power spectrum was computed \citep{Frandsen}.

\begin{figure}[!t]
\centering
\includegraphics*[scale=0.6]{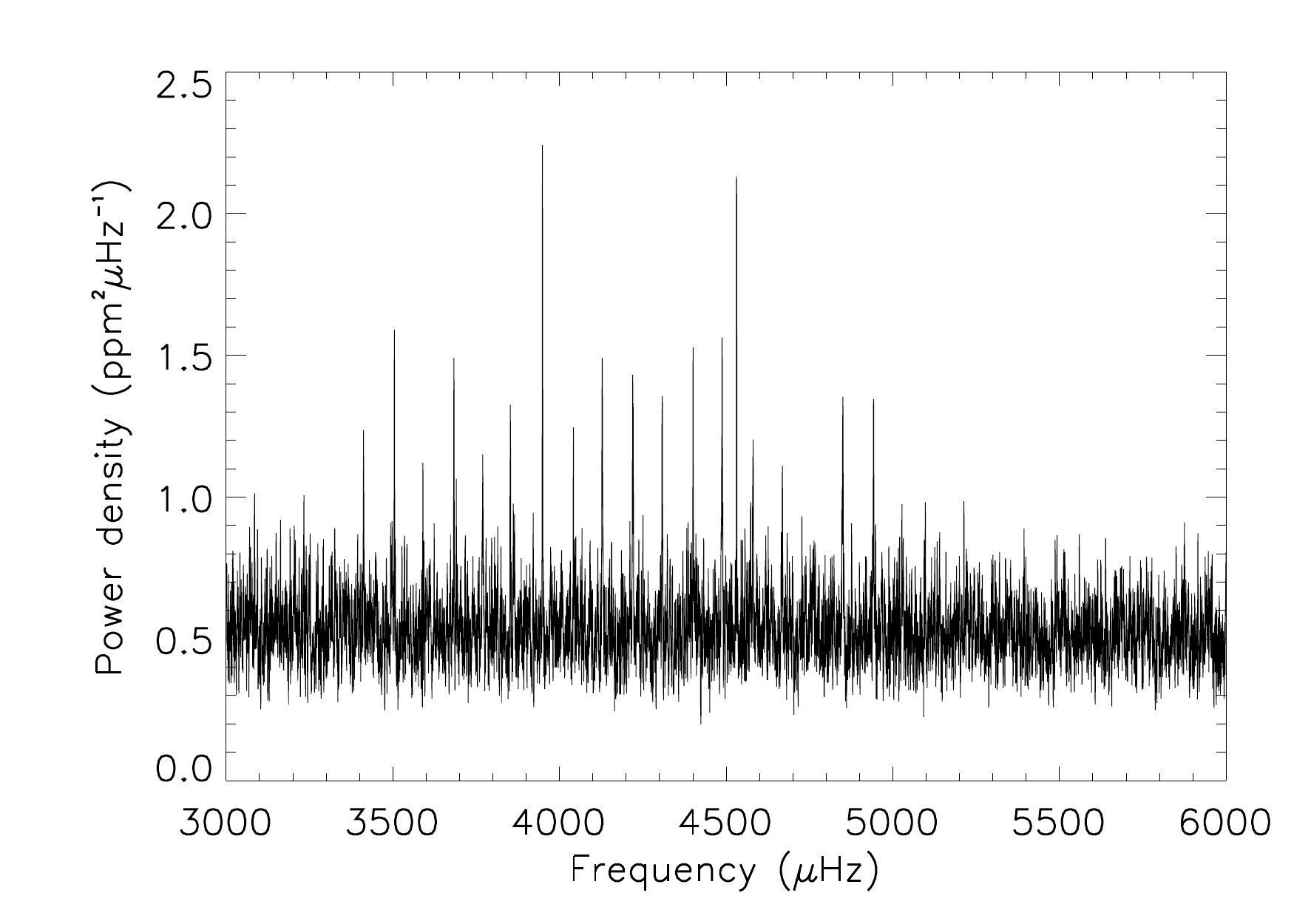}
\includegraphics*[scale=0.6]{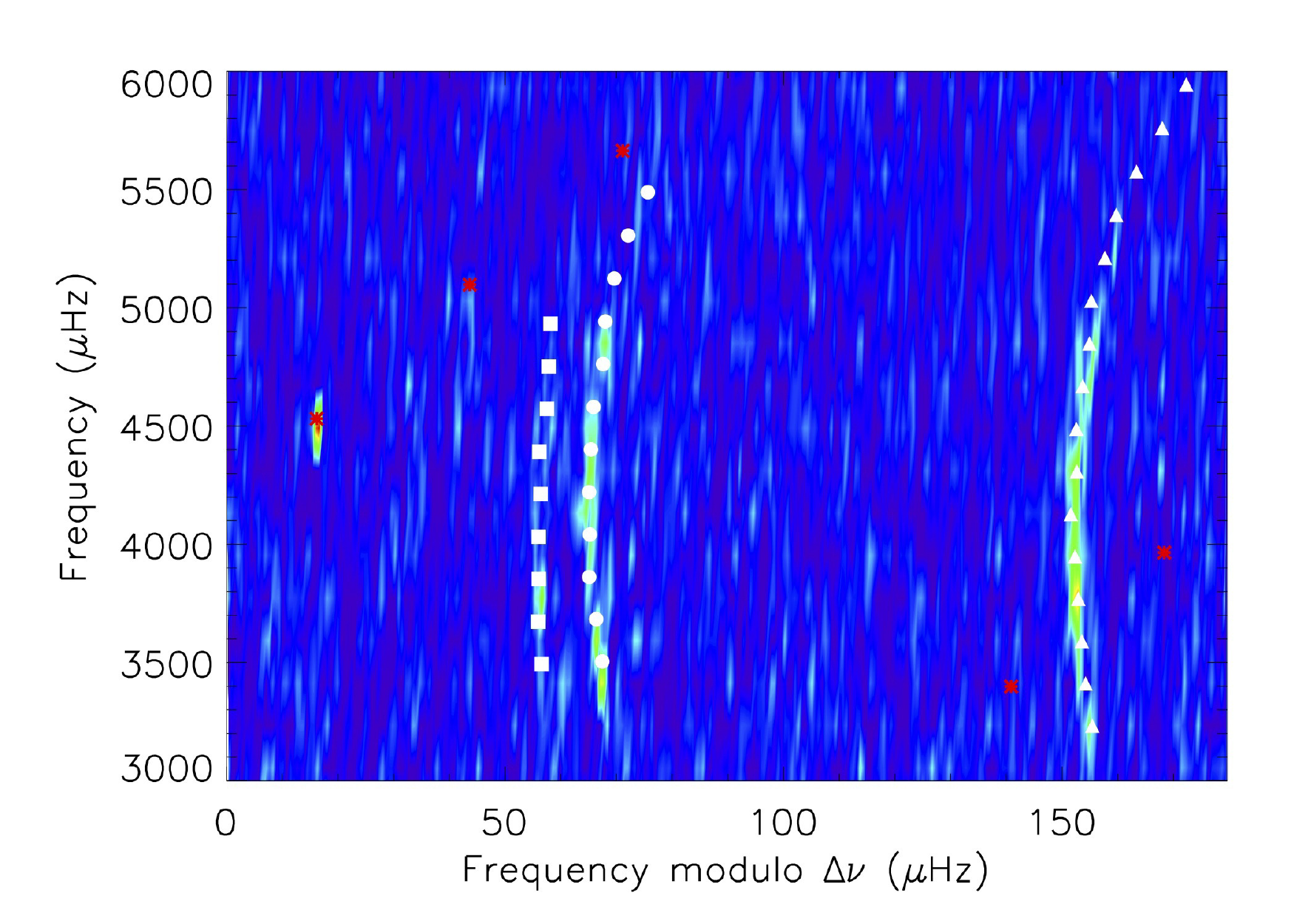}
\caption{\small Top panel: Frequency-power spectrum of the flux time series of Kepler-444 over the frequency range occupied by the solar-like oscillations. A boxcar filter of width $1.5\:{\rm \mu Hz}$ has been applied to enhance p-mode visibility. Bottom panel: Power spectrum of the {\it Kepler} light curve in \'echelle format. This is the graphical equivalent to slicing the spectrum into segments of length $\Delta \nu\!=\!179.64\:{\rm \mu Hz}$ and stacking them one on top of the other. Note that, in order to center the power ridges on the diagram, frequencies have been shifted sideways by subtracting a fixed reference of $25.5\:{\rm \mu Hz}$ (i.e., $2000\:{\rm \mu Hz} \bmod{\Delta\nu}$). White symbols represent the observed oscillation frequencies. Symbol shapes indicate mode degree: $l\!=\!0$ (circles), $l\!=\!1$ (triangles), and $l\!=\!2$ (squares). Red asterisks mark the harmonics of the inverse of the {\it Kepler} long-cadence period ($\Delta t\!\sim\!30\:{\rm min}$), not all of which are present in these data. These artifacts appear in the power spectra of {\it Kepler} short-cadence time series.\label{fig:echelle}}
\end{figure}

Mode frequencies were extracted from the power spectrum using a Markov chain Monte Carlo (MCMC) fitting procedure \citep{HandCamp,CampantePhD}. Mode frequency posterior probability distributions are typically well described by a normal distribution, so we report the median of the distribution together with the standard deviation representing the $68.3\,\%$ credible region. Mode frequencies were corrected to account for the Doppler shift due to the line-of-sight motion of the star \citep{los}. The observed oscillation frequencies (both corrected and uncorrected) are listed in Table \ref{tb:obsfreqs}. Figure \ref{fig:echelle} shows the power spectrum of the flux time series of Kepler-444.

Stellar properties were determined using three different techniques to model the oscillation frequencies extracted from the data. The first method relies on a dense grid of stellar models computed with the GARching STellar Evolution Code \citep[GARSTEC;][]{GARSTEC} including the effects of microscopic diffusion, and on theoretical frequencies calculated using the Aarhus aDIabatic PuLSation code \citep[ADIPLS;][]{ADIPLS}. The results were obtained implementing a Bayesian scheme that uses the spectroscopic constraints and frequency ratios as the parameters in the fit \citetext{V.~Silva Aguirre et al., submitted}, the latter being almost insensitive to the surface effects in solar-like oscillators \citep{Roxburgh,VSA11}. Central values are given as the estimates of the stellar properties obtained in this manner. We also computed models using the ASTEC and YREC codes. In these cases, the fit was made to the individual frequencies after correcting for the surface effect with, respectively, an appropriately scaled version of the observed solar surface correction \citep{JCD12} and a solar-type correction as described in \citet{Kepler36}. The stellar properties derived using the three techniques described above are consistent within the returned formal errors. Therefore, we added in quadrature the difference in central values of each property to the formal uncertainties determined from the GARSTEC Bayesian scheme as a measurement of the systematic spread arising from different codes and fitting techniques. In Table \ref{tb:properties}, we provide a precise estimate of the stellar age, $t$, from detailed frequency modeling. Values for the remaining fundamental stellar properties are consistent, within errors, with those obtained from grid-based modeling. In particular, no gain in precision was obtained for the stellar radius. On the other hand, the precision on the stellar mass is improved by nearly a factor of two, from $5.7\,\%$ to $3.2\,\%$.

\section{Characterization of the planetary system}\label{sec:planetary}

\subsection{Transit analysis}\label{sec:transit}
We investigated the planetary and orbital properties of the Kepler-444 system using long-cadence {\it Kepler} data \cite[$\Delta t\!\sim\!30\:{\rm min}$;][]{LCdata}. These data virtually span the entire duration of the nominal mission, with approximately four years of nearly continuous coverage. In order to limit biases induced by stellar variability and instrumental systematics, we high-pass filtered the data using a second-order Savitzky--Golay filter with a 2-day window. Data points in transit were assigned null weight in the filter to avoid diluting the transit signals. The filtering was performed independently for each quarter and the data were combined after normalizing the individual quarters to their median value.

\begin{figure}[!t]
\centering
\includegraphics[width=0.7\linewidth]{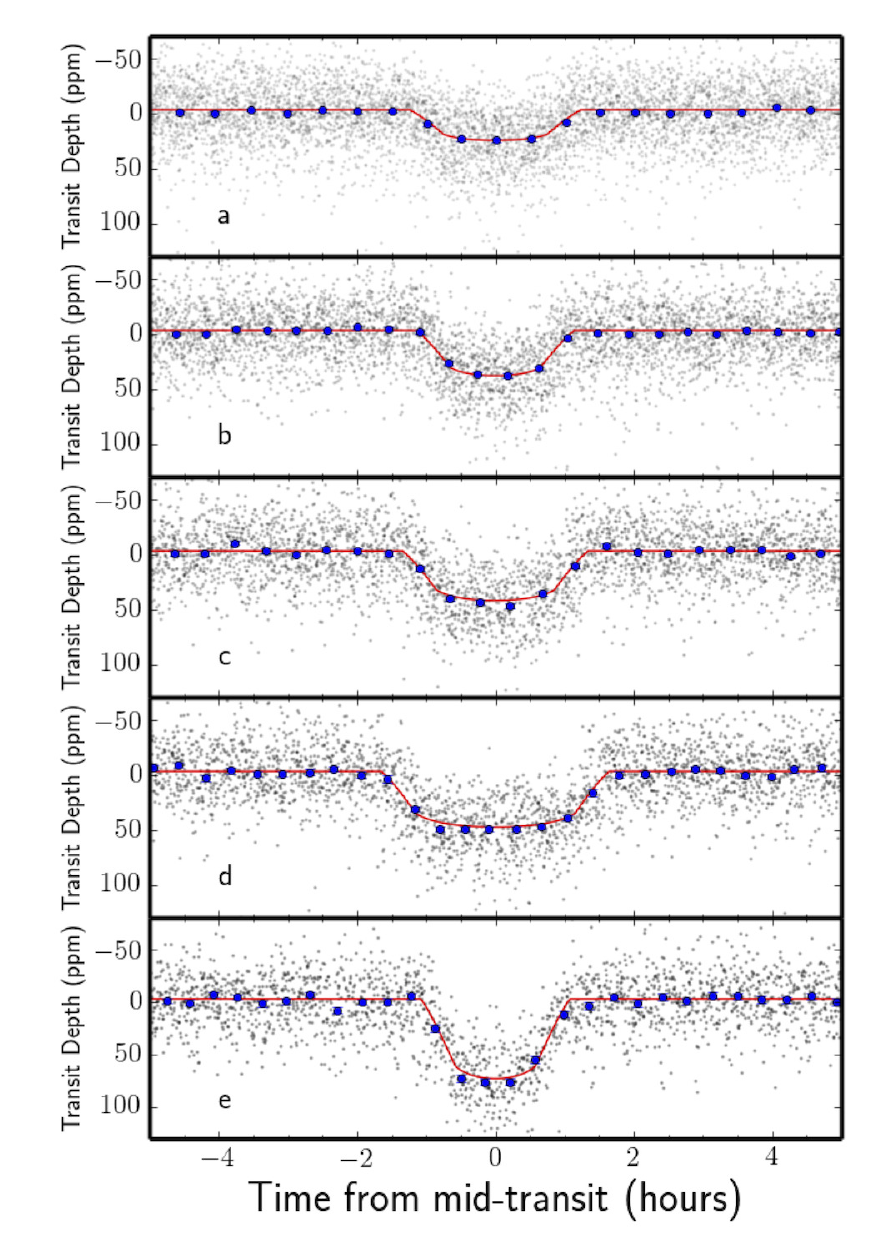}
\caption{\small Transit light curves for the five planets orbiting Kepler-444. From panels {\bf a} to {\bf e}: Transits of planets Kepler-444b, Kepler-444c, Kepler-444d, Kepler-444e, and Kepler-444f, respectively. The photometric light curves have been phase-folded on the orbital period of the planets to show the observed data as a function of orbital phase. Individual data points are shown as gray dots. Blue dots correspond to a binning of individual data points, shown only for clarity. The magnitude of the associated error bars is then given by the standard deviation of the data making up each bin divided by the square root of the number of points in the bin. These error bars are comparable in size to the blue dots. The best-fitting transit model, based on the maximum a posteriori parameter estimates, is shown as a red line.\label{fig:transit}}
\end{figure}

We measured the properties of the five planets in a similar manner to that described in \citet{Rowe14}, who used the same transiting-planet model. The model includes all five planets, with transit shapes described by the analytic model of \citet{MandelAgol02} using a quadratic limb-darkening law. The transit model computes a full orbital solution for each of the planets, which allowed for eccentricity to vary. We fixed the dilution of the light curve at $3.94\,\%$ based on our adaptive optics observations. Marginal transit-timing variations (TTVs) have been detected for the outermost planetary pair. Their amplitudes are, nevertheless, very small and do not affect the measured planetary and orbital parameters. The potential detection of TTVs in this system will be addressed in future work. Free parameters in the fit include the mean stellar density, $\langle\rho\rangle$, the two limb-darkening parameters, $\gamma_1$ and $\gamma_2$, and for each planet the time at the midpoint of the first transit, $T_0$, the orbital period, $P$, the planet-to-star radius ratio, $R_{\rm p}/R_\star$, the impact parameter, $b$, and the two eccentricity vectors $e\sin\omega$ and $e\cos\omega$, where $e$ is the orbital eccentricity and $\omega$ is the argument of periastron.

The data were modeled using an affine-invariant ensemble MCMC algorithm that utilizes multiple chains to decrease autocorrelation time \citep{GoodmanWeare,Foreman-Mackey}. The asteroseismically derived mean stellar density was used as a strong prior in the transit model, while the limb-darkening coefficients were constrained by a prior of width $0.1$ and mean derived from the model limb-darkening coefficients for the {\it Kepler} bandpass \citep{Claret11}. Furthermore, both the mid-transit time and orbital period were assigned flat priors, a uniform prior in the range $[0,1]$ was adopted for the planet-to-star radius ratio, and the prior for the impact parameter was set to a uniform distribution in the range $[0,1\!+\!R_{\rm p}/R_\star]$ to allow for grazing transits. The eccentricity vectors were assigned uniform priors in the range $[-1,1]$, although we have included an $1/e$ correction which has the effect of enforcing a uniform prior in eccentricity \citep{Eastman13}. 

We used a Gaussian likelihood function in our MCMC analysis which takes the form of a chi-squared log-likelihood \citep{Quintana14}. We work in log-likelihood space since it enhances numerical stability. We ran the MCMC with 600 chains and $30{,}000$ jumps of each chain, but discarded the first $10{,}000$ jumps as burn-in. The chains were all well mixed and converged upon a unimodal posterior distribution in each model parameter. In Table \ref{tb:planets}, we quote the median and associated $68.3\,\%$ credible region for all measured (and derived) planetary and orbital parameters. The best-fitting model is shown, plotted over the phase-folded transit data, in Fig.~\ref{fig:transit}. The best-fitting model is based on the maximum a posteriori parameter estimates.

\subsection{System validation}\label{sec:validation}
The five planet candidates associated with Kepler-444 constitute a true five-planet system orbiting the target star. In what follows we present several lines of evidence that support this conclusion.

\begin{figure}[!t]
\centering
\plottwo{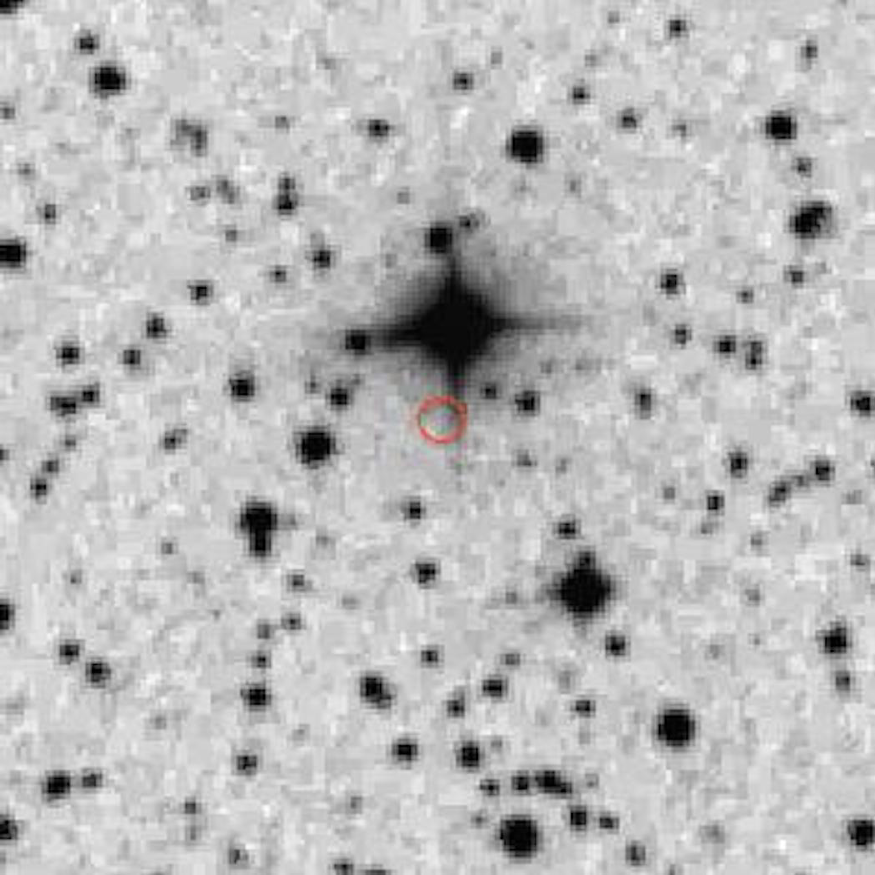}{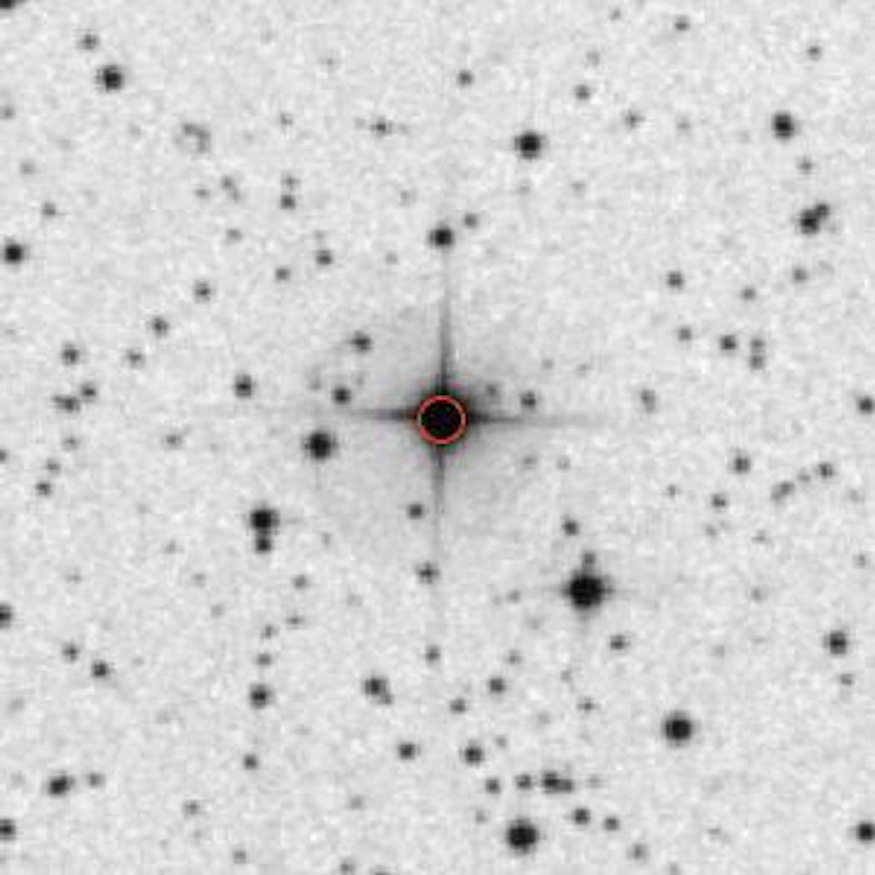}
\caption{\small Digitized Sky Survey POSS-I (left-hand panel) and POSS-II (right-hand panel) images of the region where Kepler-444 currently is. Open circles at the center of both images mark the target star's current position (J2000.0). Both images have a size of $4\!\times\!4\:{\rm arcmin}$ and were taken in the red.\label{fig:poss}}
\end{figure}

We start by invoking statistical arguments to exclude the scenario in which one or more transit-like signals are caused by chance-alignment blends such as background eclipsing binaries. Background eclipsing binaries are the primary source for the occurrence of false positives among {\it Kepler} planet candidates. Therefore, it is reasonable to assume that false positives are randomly distributed among {\it Kepler} target stars. Moreover, the number of {\it Kepler} planet candidates in multiple systems is considerably larger than what would be expected if these were assigned randomly to target stars. These considerations prompted \citet{Lissauer12} to devise a statistical approach for the validation of planet candidates in multiple systems. For a system with five planet candidates, the only plausible false-positive configuration would be that of four planets and one background eclipsing binary \citep{Lissauer12}. Chances of that happening are of $0.07\,\%$, if we assume that the fraction of candidates that are planets (viz., the fidelity of the sample) takes the rather realistic value of 0.9. We thus exclude the possibility of a background eclipsing binary at the $99.9\,\%$ level. The above statistical framework has been recently refined by using a larger, more uniform, and better vetted set of planet candidates \citep{Lissauer14}. Once again, the only plausible false-positive configuration continues to be that of four planets and one background eclipsing binary. Assuming a fidelity of the sample of single-planet candidates (and not the overall fidelity of the sample as before) of 0.9, then the chances of that happening are of $0.10\,\%$, and we are still able to exclude the possibility of occurrence of false positives at the $99.9\,\%$ level. Given the target star's high proper motion, a sanity check would be to look at Digitized Sky Survey\footnote{\url{https://archive.stsci.edu/dss/}} POSS-I (epoch 1945--58) and POSS-II (epoch 1984--99) images of the region where Kepler-444 currently is and see whether there are any background stars (Fig.~\ref{fig:poss}). We can see that there are no background stars in the above region down to the confusion limit of the images ($V\!\sim\!22$).

We now invoke the non-randomness of the observed multi-resonant chain \citep{Fabrycky12} to assert that all five planets form a single system orbiting the same star. Theoretical models of planet formation and migration suggest that there may be an excess of exoplanet pairs near mean-motion resonances \citep[MMRs;][]{Marcy01,Terquem07}. Based on a statistical sample of 408 {\it Kepler} planet candidates in multiple systems, it has been shown that the number of planetary pairs in or near MMRs exceeds that of a random distribution \citep{Lissauer11}. Moreover, the same authors found the distribution of candidate period ratios to exhibit prominent spikes near strong MMRs, implying that candidates with such period ratios are likely to be true planets. This is the case of all adjacent pairs in the Kepler-444 system, whose period ratios are just wide of first-order MMRs, being nearly commensurate. By extension, the observed multi-resonant chain provides strong evidence that all five planets form a single system. We next quantify this assertion. For each adjacent pair, we computed the variable $\zeta_{1,1}$, which measures the scaled difference between the observed period ratio and the first-order mean-motion resonance in its neighborhood \citep{Fabrycky12}. The null hypothesis being that near-resonant locations are not preferred, as would be the case if planet pairs did not orbit the same star, means that the sum $\sum |\zeta_{1,1}|$ over all adjacent pairs approximately follows an Irwin--Hall distribution\footnote{The Irwin--Hall distribution is the probability distribution of a random variable defined as the sum of a number of independent random variables, each having a uniform distribution.}. The probability that $\sum |\zeta_{1,1}|$ takes a value less than or equal to the one observed is then of about $7\,\%$. We thus reject the null hypothesis at the $90\,\%$ level. This not only means that the observed multi-resonant chain is likely not just a product of chance, but also that the resonances played a role in shaping the system's architecture. We take this as a strong indication that all five planets form a single system orbiting the same star.

We further mention stability as a boost in our confidence that these are in fact genuine planets in a multiple system. In tightly packed high-multiplicity systems, most configurations sharing the same number of planets randomly spread over the same range in period are likely to be unstable \citep{Lissauer11,Lissauer12}. Therefore, finding evidence of the system's stability will aid in its validation. We computed the nominal dynamical separation, $\Delta$, between each adjacent pair having made use of a simple power-law mass-radius relationship \citep{Lissauer11,Fabrycky12}. All dynamical separations satisfy the stability condition for two-planet systems (i.e., $\Delta\!>\!2\sqrt{3}$) and are thus Hill stable \citep{Gladman93}. Furthermore, taking each adjacent set of three planets, the $\Delta$ separations of the inner ($\Delta_{\rm inner}$) and outer ($\Delta_{\rm outer}$) pairs satisfy the heuristic stability condition $\Delta_{\rm inner}+\Delta_{\rm outer}\!>\!18$ \citep{Lissauer11}. This is taken as plausible evidence of the system's stability. Moreover, by enforcing marginal dynamical stability one requires that the five orbiting bodies are substellar in mass, thus validating their planetary nature. In this regard, we should add that determining the planetary masses from radial velocity measurements would be extremely challenging, the predicted semi-amplitude being $K\!\approx\!0.3\:{\rm m\,s^{-1}}$ even with all five planets contributing.

Finally, we establish that the planets transit the target star and not one of its M-dwarf companions. Measurements of shifts in the {\it Kepler} pixel-mask photocenter during a transit have been used in previous studies to help identify transit sources that are separated from the target stars. However, for heavily saturated target stars such as Kepler-444, a centroid analysis is highly unreliable \citep{Bryson13}. Therefore, we instead followed a different approach and tested the dynamical stability of the five planets if they were to all orbit either of the M dwarfs. Although the stellar parameters of the brighter M-dwarf companion are better constrained, both M dwarfs have similar properties ($T_{\rm eff}\!\sim\!3500\:{\rm K}$ and $\log g\!\sim\!5\:{\rm dex}$; Sect.~\ref{sec:pair}), and so we adopted these values and assumed that each contributes the same amount of dilution ($1.97\,\%$). We thus need only consider the case of planets orbiting one of these M dwarfs, since they are, for stability purposes, essentially the same. In this scenario, the M dwarf is $4.22$ {\it Kepler} magnitudes fainter than the target K star, and by interpolating over Dartmouth \citep{Dotter08} stellar evolution isochrones (assuming primary and secondaries co-evolved), we derived a mass of $0.37\,{\rm M}_\sun$ and a radius of $0.36\,{\rm R}_\sun$ for the M dwarf. The semi-major axes of the planets were computed via Kepler's law using the precise orbital period measurements. The planetary radii were computed from $R_{\rm p}/R_\star$, assuming a fixed dilution of $96.06\,\%$ from the target K star and $1.97\,\%$ from the other M dwarf, yielding (from the innermost to the outermost planet): $1.33$, $1.64$, $1.75$, $1.80$, and $2.45\,R_\earth$. Because the planetary masses are unknown (regardless of which star they orbit), we estimated their masses for an exhaustive range of compositional schemes using mass-radius relations derived from theoretical thermal evolution models \citep{Fortney07,Lopez12}. We examined the dynamical stability for planets composed of different mixtures of ice (low density, less massive planets), as well as of silicate rock and iron (high density, more massive planets). We also included $1\,\%$ (by mass) H/He gas envelopes, certainly pertinent to planets larger than about $1.7$ to $2\,R_\earth$ as these may be mini-Neptunes rather than super-Earths with solid surfaces, although planets this close to their star (within $0.06\:{\rm AU}$ in this scenario) are highly irradiated and would thus be vulnerable to atmospheric escape \citep{Lopez12}. For all compositional schemes tested, the planets became unstable within $10^2$ to several $10^3$ years, even in the extremely unlikely case of pure ice planets. These results decisively support the competing scenario that the planets orbit the target K dwarf. We thus conclude that the five planet candidates associated with Kepler-444 constitute a true five-planet system orbiting the target K star.

\section{Discussion and conclusions}\label{sec:discussion}
The precision with which we measured the planetary radii varies between $2.9\,\%$ and $5.5\,\%$. Kepler-444b is the innermost and smallest planet (within $2\sigma$ of the size of Mercury). Its radius was measured with a precision of $\sim\!100\:{\rm km}$. All five planets are sub-Earth-sized with monotonically increasing radii as a function of orbital distance: $0.403$, $0.497$, $0.530$, $0.546$, and $0.741\,R_\earth$. Kepler-444c, Kepler-444d, and Kepler-444e have very similar radii, respectively within $2\sigma$, $1\sigma$, and $1\sigma$ of the size of Mars. Finally, Kepler-444f has a size intermediate to Mars and Venus. Kepler-444 thus expands the population of planets found in low-metallicity environments from the mini-Neptunes around Galactic halo's Kapteyn's star \citep{Kapteyn} down to the regime of terrestrial-size planets. Although photometry alone does not yield the masses of the planets, planetary thermal evolution models \citep{Lopez13} predict that the composition of planets with radii less than $0.8\,R_\earth$ are highly likely rocky.

The parent star, and hence the planetary system, has an age of $11.2\pm1.0\:{\rm Gyr}$ from detailed frequency modeling. Kepler-444 is slightly older than Kepler-10, the host of two rocky super-Earths \citep{Kepler10,Kepler10Dumusque}, whose age of $10.4\pm1.4\:{\rm Gyr}$ has also been determined from a detailed modeling of the oscillation frequencies \citep{Kepler10age}. The precision with which the age of Kepler-444 has been determined from asteroseismology ($\sim\!9\,\%$) is an impressive technical achievement that was only made possible due to the extended and high-quality photometry provided by the {\it Kepler} mission. We have thus attained the level of precision expected for ESA's future {\it PLATO} mission, which has the science goal of providing stellar ages to $10\,\%$ precision as a key to exoplanet parameter accuracy \citep{PLATO}. The estimated stellar age is commensurate with the ages of thick-disk field stars, which are older than about $10$--$11\:{\rm Gyr}$ \citep{Reddy06}. It is also in general good agreement with several studies of the Arcturus stream based on different selections of stream members \citep{Navarro04,Helmi06,Williams09,Ramya12}. In \citet{Ramya12}, stellar ages were determined for a selection of stream members based on their location in a color-magnitude diagram. Ages of stream members were seen to vary between 10 and $14\:{\rm Gyr}$. An individual age estimate for Kepler-444 was given by $13.4^{+3.2}_{-0.1}\:{\rm Gyr}$, thus being significantly greater than the asteroseismic age estimate. Kepler-444 is the oldest known system of terrestrial-size planets. We thus show that Earth-size planets have formed throughout most of the Universe's $13.8$-billion-year history, providing scope for the existence of ancient life in the Galaxy. Remarkably, by the time Earth formed, this star and its Earth-size companions were already older than our planet is today. 

\begin{figure}[!t]
\centering
\includegraphics[width=\linewidth]{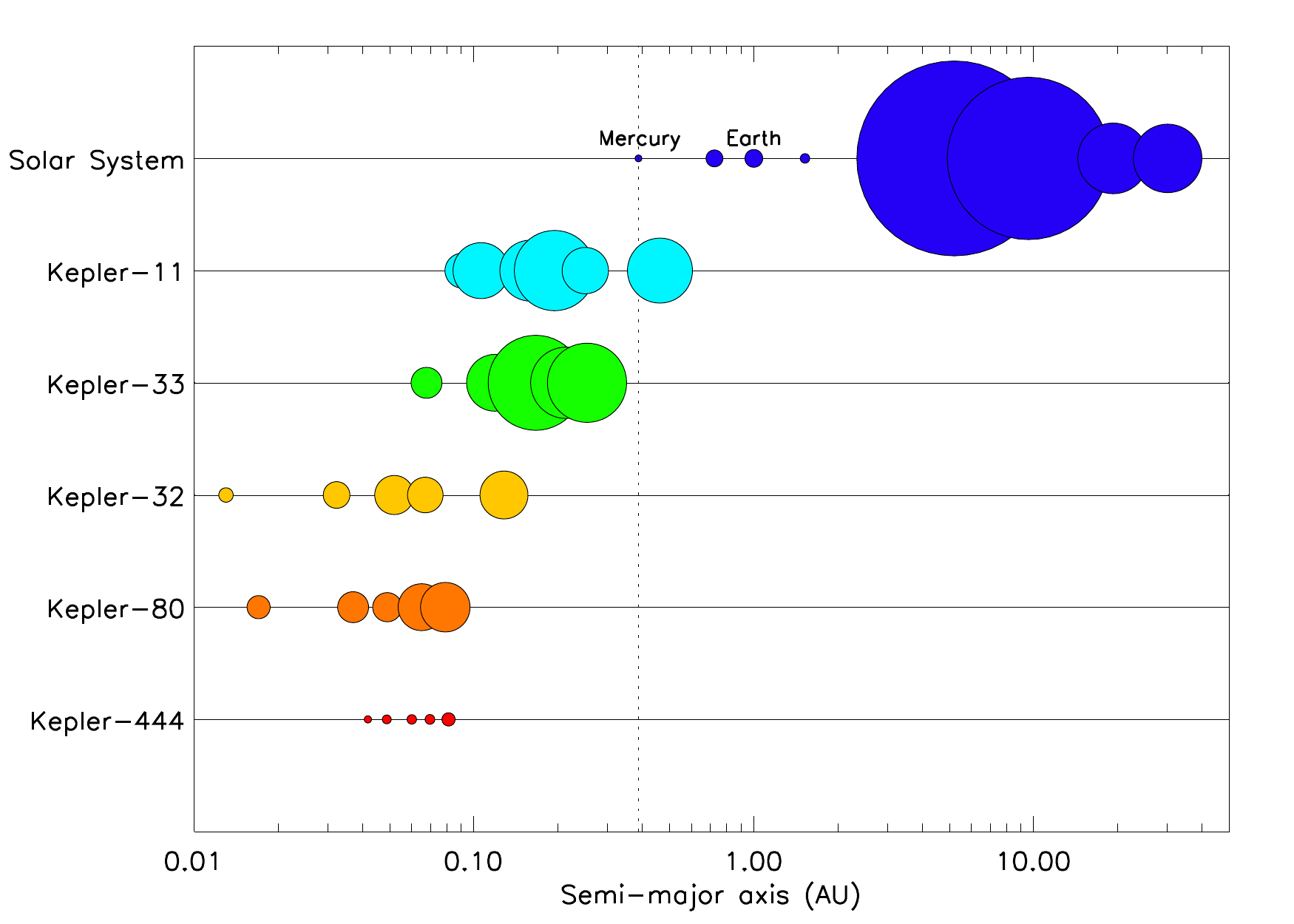}
\caption{\small Semi-major axes of planets belonging to the highly-compact multiple-planet systems Kepler-444, Kepler-11, Kepler-32, Kepler-33, and Kepler-80. Semi-major axes of planets in the Solar System are shown for comparison. The vertical dotted line marks the semi-major axis of Mercury. Symbol size is proportional to planetary radius. Note that all planets in the Kepler-444 system are interior to the orbit of the innermost planet in the Kepler-11 system, the prototype of this class of highly-compact multiple-planet systems.\label{fig:compact}}
\end{figure}

This system is highly compact, both in terms of its architecture and in a dynamical sense. All planets orbit the parent star in less than 10 days, or within $0.08\:{\rm AU}$, roughly one-fifth the size of Mercury's orbit. These orbits are well interior to the inner edge of the system's habitable zone, which lies $0.47\:{\rm AU}$ from the parent star if we consider the rather optimistic `Recent Venus' limit of \citet{Kopparapu}. Also, the orbits of the planets are consistent with having zero mutual inclinations, an indication of near-coplanarity. Furthermore, all adjacent planet pairs are close to being in exact orbital resonances, with period ratios no more than $\sim\!2\,\%$ in excess of strong $5\colon\!4$, $4\colon\!3$, $5\colon\!4$, and $5\colon\!4$ first-order MMRs as one moves toward the outermost pair. This latter feature could potentially facilitate the determination of the planetary masses via TTVs \citep[e.g.,][]{WuLithwick}. Other highly-compact multiple-planet systems -- characterized by a concentration of dynamically packed planets near $0.1\:{\rm AU}$ -- include, e.g., Kepler-11 \citep{Kepler-11}, Kepler-32 \citep{Kepler-32}, Kepler-33 \citep{Lissauer12}, and Kepler-80 \citep{Rowe14}, all with at least five planets (Fig.~\ref{fig:compact}). Kepler-444 is thus a member of a distinct population of highly-compact planetary systems, which make up only $\sim\!1\,\%$ of all {\it Kepler} targets with planetary candidates \citep{Lissauer11}.

The proximity of each planet to a strong, first-order MMR indicates that this system evolved dynamically after the formation of the planets. A likely mechanism to produce such a configuration is convergent inward migration within a gaseous or planetesimal disk. Since these planets are all sub-Earth-sized and likely of similar composition \citep{WeissMarcy}, the monotonic increase in planetary size as a function of orbital distance would imply that the masses of the planets increase outward. This would provide a means for convergent migration, since the migration rate is expected to scale as planet mass \citep{KleyNelson}. The subsequent damping of orbital eccentricities as a result of tidal evolution would tend to spread the orbits, pushing them wide of exact commensurability as observed.

In addition to the dynamical evolution, the chemical environment is thought to play a decisive role in the formation of systems of terrestrial-size planets such as Kepler-444. While gas-giant planets appear to form preferentially around metal-rich stars, small planets (with radii less than $4\,R_\earth$) can form under a wide range of metallicities \citep{Sousa-11,Buchhave-12,Buchhave14}. This could mean that the process of formation of small, including Earth-size, planets is less selective than that of gas giants, with the former likely starting to form at an earlier epoch in the Universe's history when metals were far less abundant \citep{Fischer-12}. There is growing evidence that the critical elements for planet formation in iron-poor environments are $\alpha$-process elements \citep{Cochran,Brugamyer-11,Adibekyan-12b}. In particular, $\alpha$ elements comprise the bulk of the material that constitutes rocky, Earth-size planets \citep{Valencia07,Valencia10}. Stars belonging to the thick disk are overabundant in $\alpha$ elements compared to thin-disk stars in the low-metallicity regime \citep{Reddy06}, which may explain the greater planet incidence among thick-disk stars for metallicities below half that of the Sun \citep{Adibekyan-12a}. Similarly favorable conditions to planet formation in iron-poor environments seem to be associated with a fraction of the halo stellar population, namely, the so-called high-$\alpha$ stars \citep{Nissen10}. Thus, thick-disk and high-$\alpha$ halo stars were likely hosts to the first Galactic planets. The discovery of an ancient system of terrestrial-size planets around the thick-disk star Kepler-444 confirms that the first planets formed very early in the history of the Galaxy, thus helping to pinpoint the beginning of the {\it era of planet formation}.

The first discoveries of exoplanets around Sun-like stars \citep{firstexo,MarcyButler} have fueled efforts to find ever smaller worlds evocative of Earth and other terrestrial planets in the Solar System. From the first rocky exoplanets \citep{CoRoT-7,Kepler10} to the discovery of an Earth-size planet orbiting another star in its habitable zone \citep{Quintana14}, we are now getting first glimpses of the variety of Galactic environments conducive to the formation of these small worlds. As a result, the path toward a more complete understanding of early planet formation in the Galaxy starts unfolding before us.

\acknowledgments
Funding for the {\it Kepler} mission is provided by NASA's Science Mission Directorate. The authors wish to thank the entire {\it Kepler} team, without whom these results would not be possible. We thank David W.~Latham for helpful comments on the manuscript. Funding for the Stellar Astrophysics Centre is provided by The Danish National Research Foundation (grant DNRF106). The research is supported by the ASTERISK project (ASTERoseismic Investigations with SONG and {\it Kepler}) funded by the European Research Council (grant agreement no.~267864). The research leading to the presented results has received funding from the European Research Council under the European Community's Seventh Framework Programme (FP7/2007-2013)/ERC grant agreement no.~338251 (StellarAges). This research made use of \texttt{APLpy}, an open-source plotting package for \texttt{Python} hosted at \url{http://aplpy.github.com}. T.L.C., G.R.D., W.J.C., R.H., A.M., and Y.P.E.~acknowledge the support of the UK Science and Technology Facilities Council (STFC). T.B.~was supported by a NASA Keck PI Data Award, administered by the NASA Exoplanet Science Institute. Some of the data presented herein were obtained at the W.~M.~Keck Observatory from telescope time allocated to the National Aeronautics and Space Administration through the agency's scientific partnership with the California Institute of Technology and the University of California. The Observatory was made possible by the generous financial support of the W.~M.~Keck Foundation. The authors wish to recognize and acknowledge the very significant cultural role and reverence that the summit of Mauna Kea has always had within the indigenous Hawaiian community. We are most fortunate to have the opportunity to conduct observations from this mountain. D.H.~and E.V.Q.~acknowledge support by an appointment to the NASA Postdoctoral Program at Ames Research Center administered by Oak Ridge Associated Universities. D.H.~also acknowledges NASA grant NNX14AB92G issued through the {\it Kepler} Participating Scientist Program and support by the Australian Research Council's Discovery Projects funding scheme (project no.~DE40101364). V.Zh.A., N.C.S., and S.G.S.~acknowledge support from the European Research Council/European Community under FP7 through Starting grant agreement no.~239953, and from FCT (Portugal) through FEDER funds in program COMPETE, as well as through national funds in the form of grants RECI/FIS-AST/0176/2012 (FCOMP-01-0124-FEDER-027493) and RECI/FIS-AST/0163/2012 (FCOMP-01-0124-FEDER-027492). V.Zh.A.~and S.G.S.~also acknowledge grants SFRH/BPD/70574/2010 and SFRH/BPD/47611/2008 from FCT (Portugal), respectively. The Robo-AO system is supported by collaborating partner institutions, the California Institute of Technology and the Inter-University Centre for Astronomy and Astrophysics, and by the National Science Foundation under grant nos.~AST-0906060 and AST-0960343, by the Mount Cuba Astronomical Foundation, and by a gift from Samuel Oschin. C.B.~acknowledges support from the Alfred P.~Sloan Foundation. S.B.~acknowledges NSF grant AST-1105930 and NASA grant NNX13AE70G. T.S.M.~acknowledges NASA grant NNX13AE91G. C.K.~acknowledges support from the Villum Foundation. N.C.S.~also acknowledges support in the form of contract reference IF/00169/2012 funded by FCT/MCTES (Portugal) and POPH/FSE (EC).

{\it Facilities:} \facility{Kepler}, \facility{Keck:I (HIRES)}, \facility{Keck:II (NIRC2)}, \facility{PO:1.5m (Robo-AO)}

\bibliographystyle{apj}
\bibliography{biblio}

\begin{deluxetable}{rc}
\tablecolumns{2}
\tablewidth{0pc}
\tablecaption{Atmospheric parameters and elemental abundances.\label{tb:spec}}
\tablehead{\colhead{Parameter} & \colhead{Value}}
\startdata
$T_{\rm eff}$ (K)\tablenotemark{a} & $5046\pm74\,(44)$ \\
$\log g_{\rm spec}$ (dex) & $4.595\pm0.060$ \\
$[{\rm Fe}/{\rm H}]$ (dex)\tablenotemark{a} & $-0.55\pm0.07\,(0.03)$ \\
$[{\rm Si}/{\rm H}]$ (dex) & $-0.28\pm0.02$ \\
$[{\rm Ti}/{\rm H}]$ (dex) & $-0.30\pm0.05$ \\
\enddata
\tablenotetext{a}{\small Contributions of $59\:{\rm K}$ in $T_{\rm eff}$ and $0.062\:{\rm dex}$ in $[{\rm Fe}/{\rm H}]$ were added in quadrature to the formal uncertainties shown in parentheses.}
\end{deluxetable}

\begin{deluxetable}{rc}
\tablecolumns{2}
\tablewidth{0pc}
\tablecaption{Fundamental stellar properties.\label{tb:properties}}
\tablehead{\colhead{Parameter} & \colhead{Value}}
\startdata
$M/{\rm M}_\sun$ & $0.758\pm0.043$ \\
$R/{\rm R}_\sun$ & $0.752\pm0.014$ \\
$\log g_{\rm seis}$ (dex) & $4.5625\pm0.0095$ \\
$\langle\rho\rangle$ (${\rm g\,cm^{-3}}$) & $2.493\pm0.028$ \\
$t$ (Gyr) & $11.23^{+0.91}_{-0.99}$ \\
\enddata
\tablecomments{\small Stellar age was determined from detailed frequency modeling. All remaining properties were determined from grid-based modeling.}
\end{deluxetable}

\begin{deluxetable}{cccccc}
\tablecolumns{6}
\tablewidth{0pc}
\tablecaption{Observed oscillation frequencies.\label{tb:obsfreqs}}
\tablehead{
\colhead{} & \multicolumn{2}{c}{Uncorrected} & \colhead{} & 
\multicolumn{2}{c}{Corrected} \\
\cline{2-3} \cline{5-6} \\
\colhead{$l$} & \colhead{Frequency ($\rm{\mu Hz}$)} & \colhead{Uncertainty ($\rm{\mu Hz}$)} & \colhead{} &
\colhead{Frequency ($\rm{\mu Hz}$)\tablenotemark{a}} & \colhead{Uncertainty ($\rm{\mu Hz}$)\tablenotemark{a}}}
\startdata
0&3504.57&0.34&&3503.16&0.30\\
0&3683.12&0.15&&3681.62&0.19\\
0&3861.50&0.32&&3859.95&0.32\\
0&4041.25&0.15&&4039.61&0.19\\
0&4220.76&0.61&&4218.94&0.65\\
0&4400.74&0.26&&4398.93&0.27\\
0&4580.84&0.32&&4578.97&0.33\\
0&4762.20&0.99&&4760.56&0.97\\
0&4942.24&0.54&&4940.25&0.48\\
0&5123.48&0.92&&5121.34&0.91\\
0&5305.62&0.98&&5303.53&0.95\\
0&5488.79&1.05&&5486.60&1.06\\
\tableline
1&3233.26&0.70&&3231.90&0.66\\
1&3411.70&0.53&&3410.31&0.52\\
1&3590.67&0.89&&3589.17&0.79\\    
1&3769.67&0.26&&3768.14&0.27\\
1&3948.73&0.13&&3947.13&0.12\\
1&4127.62&0.10&&4125.94&0.07\\
1&4308.32&0.25&&4306.58&0.28\\    
1&4487.89&0.36&&4486.01&0.40\\
1&4668.67&0.49&&4666.76&0.46\\
1&4849.51&0.38&&4847.53&0.35\\
1&5029.49&1.05&&5027.48&1.09\\    
1&5211.57&1.05&&5209.39&1.06\\
1&5393.24&0.86&&5391.18&0.87\\
1&5576.51&0.91&&5574.40&0.92\\
1&5760.77&0.77&&5758.44&0.78\\    
1&5944.74&0.94&&5942.30&0.99\\
\tableline
2&3493.67&1.03&&3492.23&0.97\\
2&3672.71&0.49&&3671.21&0.52\\
2&3852.45&0.31&&3850.81&0.27\\
2&4032.07&0.27&&4030.45&0.24\\
2&4212.04&0.76&&4210.38&0.76\\
2&4391.47&0.80&&4389.74&0.77\\
2&4572.47&0.68&&4570.62&0.66\\
2&4752.43&0.83&&4750.56&0.80\\
2&4932.40&0.91&&4930.32&0.91\\
\enddata
\tablenotetext{a}{\small The correction for the line-of-sight Doppler velocity shift modifies the frequencies and associated uncertainties by multiplying these quantities with $1\!+\!V_{\rm r}/c$, where $V_{\rm r}$ is the radial velocity and $c$ is the speed of light \citep[for further details see][]{los}.}
\end{deluxetable}

\begin{deluxetable}{cccccc}
\tablecolumns{6}
\rotate
\tablewidth{0pc}
\tablecaption{Planetary and orbital parameters.\label{tb:planets}}
\tablehead{\colhead{Parameter} & \colhead{Kepler-444b} & \colhead{Kepler-444c} & \colhead{Kepler-444d} & \colhead{Kepler-444e} & \colhead{Kepler-444f}}
\startdata
$T_0$ (${\rm BJD}\!-\!2{,}454{,}833$) & $133.2599^{+0.0018}_{-0.0018}$ & $131.5220^{+0.0013}_{-0.0013}$ & $134.7869^{+0.0015}_{-0.0015}$ & $135.0927^{+0.0018}_{-0.0018}$ & $134.8791^{+0.0011}_{-0.0011}$ \\[0.12in]

$P$ (days) & $3.6001053^{+0.0000083}_{-0.0000080}$ & $4.5458841^{+0.0000070}_{-0.0000071}$ & $6.189392^{+0.000012}_{-0.000012}$ & $7.743493^{+0.000017}_{-0.000016}$ & $9.740486^{+0.000013}_{-0.000013}$ \\[0.12in]

$R_{\rm p}/R_\star$ & $0.00491^{+0.00017}_{-0.00014}$ & $0.00605^{+0.00025}_{-0.00017}$ & $0.00644^{+0.00023}_{-0.00020}$ & $0.00664^{+0.00016}_{-0.00014}$ & $0.00903^{+0.00046}_{-0.00047}$ \\[0.12in]

$R_{\rm p}/R_\earth$ & $0.403^{+0.016}_{-0.014}$ & $0.497^{+0.021}_{-0.017}$ & $0.530^{+0.022}_{-0.019}$ & $0.546^{+0.017}_{-0.015}$ & $0.741^{+0.041}_{-0.040}$ \\[0.12in]

$b$ & $0.40^{+0.17}_{-0.25}$ & $0.42^{+0.22}_{-0.27}$ & $0.53^{+0.13}_{-0.23}$ & $0.29^{+0.16}_{-0.17}$ & $0.79^{+0.07}_{-0.13}$ \\[0.12in]

$e\sin\omega$ & $0.01^{+0.08}_{-0.12}$ & $0.18^{+0.10}_{-0.15}$ & $0.03^{+0.12}_{-0.12}$ & $-0.008^{+0.040}_{-0.090}$ & $0.09^{+0.20}_{-0.15}$ \\[0.12in]

$e\cos\omega$ & $0.00^{+0.20}_{-0.21}$ & $0.01^{+0.28}_{-0.25}$ & $0.00^{+0.21}_{-0.19}$ & $-0.01^{+0.11}_{-0.21}$ & $-0.06^{+0.19}_{-0.33}$ \\[0.12in]

$e$\tablenotemark{a} & $0.16^{+0.21}_{-0.10}$ & $0.31^{+0.12}_{-0.15}$ & $0.18^{+0.16}_{-0.12}$ & $0.10^{+0.20}_{-0.07}$ & $0.29^{+0.20}_{-0.19}$ \\[0.12in]

$a/R_\star$ & $11.951^{+0.046}_{-0.046}$ & $13.961^{+0.053}_{-0.053}$ & $17.151^{+0.066}_{-0.066}$ & $19.913^{+0.076}_{-0.076}$ & $23.205^{+0.089}_{-0.089}$ \\[0.12in]

$a$ (AU) & $0.04178^{+0.00079}_{-0.00079}$ & $0.04881^{+0.00093}_{-0.00093}$ & $0.0600^{+0.0011}_{-0.0011}$ & $0.0696^{+0.0013}_{-0.0013}$ & $0.0811^{+0.0015}_{-0.0015}$ \\[0.12in]

$i$ (deg) & $88.0^{+1.2}_{-0.6}$ & $88.2^{+1.2}_{-1.0}$ & $88.16^{+0.81}_{-0.55}$ & $89.13^{+0.54}_{-0.52}$ & $87.96^{+0.36}_{-0.31}$ \\
\enddata
\tablecomments{\small Besides the free parameters in the fit, we have also included the planet radius in Earth radii, $R_{\rm p}/R_\earth$, the eccentricity, $e$, the semi-major axis in stellar radii, $a/R_\star$, the semi-major axis, $a$, and the orbital inclination, $i$.}
\tablenotetext{a}{\small The eccentricity has a non-Gaussian posterior distribution and so the median is an overestimate of the true eccentricity.}
\end{deluxetable}

\end{document}